\documentclass[twocolumn,superscriptaddress,nofootinbib,floatfix,aps,prb,citeautoscript,longbibliography,10pt]{revtex4-2}

\usepackage[utf8]{inputenc}
\usepackage[T1]{fontenc}
\usepackage{orcidlink}
\usepackage{graphicx}
\usepackage{xcolor}
\usepackage{array}
\usepackage{makecell}
\usepackage{multirow}
\usepackage{chemformula}

\usepackage{booktabs}
\usepackage{tabularx}
\usepackage{amsmath}
\usepackage{amssymb}
\usepackage{siunitx}
\usepackage{hyperref}
\usepackage[capitalise]{cleveref}

\makeatletter
\let\pineir@mkpream\@mkpream
\let\pineir@classx\@classx
\let\pineir@insertcolumn\insert@column
\AtBeginDocument{%
  \let\@mkpream\pineir@mkpream
  \let\@classx\pineir@classx
  \let\insert@column\pineir@insertcolumn}
\makeatother

\graphicspath{{figures/}}

\usepackage{xspace}
\newcommand{\PineIR}{PineIR\xspace}
\newcommand{\PineDB}{PineIRDB\xspace}
\newcommand{\pIR}{pIR\xspace}

\newcommand{\PCQM}{PCQM4Mv2\xspace}
\newcommand{\GEOM}{GEOM-QM9\xspace}
\newcommand{\QMeS}{QMe14S\xspace}

\newcommand{\QMeSDFT}{QMe14S@DFT\xspace}

\newcommand{\wn}{\ensuremath{\,\text{cm}^{-1}}\xspace}

\begin{document}

\newcommand{\AffOne}{Theoretical Chemistry, VU University Amsterdam, The Netherlands}
\newcommand{\AffTwo}{Research Center Future Energy Materials and Systems of the University Alliance Ruhr and Interdisciplinary Centre for Advanced Materials Simulation, Ruhr University Bochum, Universitätsstraße 150, D-44801 Bochum, Germany}

\title{PineIR: 3D molecular structures from an infrared spectrum alone}

\author{Giulio Benedini\,\orcidlink{0009-0000-0943-8744}}
\affiliation{\AffOne}
\affiliation{\AffTwo}

\author{Pierre-Paul De Breuck\,\orcidlink{0000-0002-3173-2058}}
\affiliation{\AffTwo}

\author{Miguel A. L. Marques\,\orcidlink{0000-0003-0170-8222}}
\affiliation{\AffTwo}

\author{Silvana Botti\,\orcidlink{0000-0002-4920-2370}}
\email{silvana.botti@rub.de}
\affiliation{\AffTwo}

\date{\today}

\begin{abstract}
Computing the infrared (IR) spectrum of a molecule from its structure is routine. The inverse problem is not. In practice, IR spectra are used to identify the functional groups present in a molecule, while the determination of its complete structure requires additional measurements. Machine-learning models that return 3D structures from IR spectra exist, but all of them need extra input, such as Raman, ultraviolet or nuclear magnetic resonance spectra, the molecular formula, or the number of atoms. In this work we present \PineIR, a generative model that returns 3D molecular structures from the IR spectrum alone. \PineIR builds the molecule atom by atom, and each atom is associated with its contribution to the IR spectrum. This contribution is subtracted from the input spectrum before the next atom is generated. To train it, we computed with a machine-learned interatomic potential the IR spectra of 4.6 million structures of 3.1 million molecules, including conformer ensembles. On computed spectra of held-out molecules, the best-ranked of fifty candidates has the correct connectivity in 43\% of the cases, and the generated geometries reproduce the conformer and the diastereomer that produced the input spectrum. After fine-tuning on 161k spectra calculated with the hybrid functional B3LYP, the model can also be applied to spectra at this level of theory.
\end{abstract}

\maketitle

\section{Introduction}
\label{sec:intro}

Infrared (IR) spectroscopy is one of the most widely used tools to characterise molecules. It is cheap, fast and non-destructive, it works in the gas, liquid and solid phase, and its bands carry information on the molecular structure~\cite{griffithsFourierTransformInfrared2007}. However, the analysis of an IR spectrum is still mostly limited to the identification of functional groups~\cite{fineSpectralDeepLearning2020}. Experts do not determine a complete structure from IR spectroscopy alone, but combine it with nuclear magnetic resonance (NMR), mass spectrometry and chemical intuition~\cite{mirzaEndtoendMultimodalStructure2026}.

The difficulty is that the inverse problem is ill-posed~\cite{hadamard1902problemes}. Different molecules can have nearly identical spectra~\cite{chen2026speccal}, while the spectra of different conformers of the same molecule usually differ moderately, and for some flexible molecules even strongly~\cite{prachtComprehensiveAssessmentGFN2020}. Structure elucidation from IR spectra should therefore be treated as a problem of statistical inference, and not of regression. A useful model must return several candidate structures and rank them.

Three machine-learning approaches have been proposed for this problem. \textit{Retrieval methods} compare the spectrum with a reference library~\cite{lavineCriteriaComparingInfrared2020, yangCrossModalRetrievalBetween132021, rocabert-oriolsMultimodalContrastiveLearning2025, kanakalaSpectraStructureContrastive2024} and can only find molecules that the library contains. \textit{2D generative models} predict molecular graphs, or their representation as strings in the simplified molecular-input line-entry system (SMILES), directly from the spectrum~\cite{wuTransformerBasedModelsPredicting2025, mitsuishiDeepLearningAssistedNovo2025, albertsLeveragingInfraredSpectroscopy2024, alberts2025setting, zhangCompleteMolecularStructure2025, ellisDeepQLearningBasedMolecular2024}, and some of them no longer need the molecular formula~\cite{mitsuishiDeepLearningAssistedNovo2025, mick2026data, zhangCompleteMolecularStructure2025, ellisDeepQLearningBasedMolecular2024}. SMEN and SECS link retrieval and generation through a learned joint embedding of spectra and molecules: SMEN decodes SMILES strings from it~\cite{kanakalaSpectraStructureContrastive2024}, and SECS uses it to score structures evolved by a genetic algorithm from the closest library molecules~\cite{mirzaEndtoendMultimodalStructure2026}. None of these methods returns the 3D geometry. Finally, \textit{3D generative models}, built on equivariant and latent diffusion~\cite{pmlr-v162-hoogeboom22a, xu2023geoldm, vignacMiDiMixedGraph2023, reidenbachApplicationsModularCodesign2026}, have recently been conditioned on spectra. DiffSpectra generates 2D and 3D structures with a diffusion model~\cite{wangDiffSpectraMolecularStructure2025v2}. MAST (motif-augmented diffusion with search tree) extends this approach with a prior on molecular substructures and a tree search over the denoising steps~\cite{jiaMASTMotifAugmentedDiffusion2026}. IR-GeoDiff generates 3D geometries from IR spectra with a latent diffusion model~\cite{wuLatentDiffusionbased3D2026}. In all of them the number of atoms is fixed before generation, and in IR-GeoDiff also the complete molecular formula~\cite{wuLatentDiffusionbased3D2026}.

The 3D geometry is chemically relevant. The conformers of a molecule determine which reaction pathways it follows~\cite{CurtinHammettPrinciple2014} and whether it binds a target~\cite{changLigandConfigurationalEntropy2007}, and they have distinct vibrational bands~\cite{barnesInfraredCryogenicStudies1970}. The IR spectrum therefore contains information on the 3D geometry that is lost in the molecular graph.

To the best of our knowledge, no method returns a 3D structure from the IR spectrum by itself (Table~\ref{tab:comparison}). All existing methods require a second spectroscopic technique (Raman and ultraviolet for DiffSpectra and MAST, NMR for SECS), the molecular formula, or the number of atoms. The formula can usually be measured, but supplying it changes the problem. With a fixed composition, structure elucidation reduces to the ranking of isomers, and a second measurement becomes necessary to interpret an IR spectrum. A second obstacle is the lack of training data. Quantum-chemical IR spectra are expensive to calculate, and the available datasets are limited to molecules of the QM9 database, which contains small organic molecules with up to nine heavy atoms (C, N, O and F)~\cite{ramakrishnanQuantumChemistryStructures2014, krishnadasSpectralQuantumChemistry2026, kanakalaSpectraStructureContrastive2024}, or to a few hundred thousand molecules~\cite{yuanQMe14SComprehensiveEfficient, NEURIPS2024_e38e60b3, mitsuishiDeepLearningAssistedNovo2025}. Machine-learned interatomic potentials (MLIPs) now compute vibrational spectra at a small fraction of the cost of density-functional theory (DFT)~\cite{prachtEfficientCompositeInfrared2024a, kabyldaMolecularSimulationsPretrained2025, anstineAIMNet2NeuralNetwork2025, bhatiaMACE4IRmolUncertaintyawareFoundation2026}, which makes much larger datasets possible.

\begin{table*}[htb!]
\caption{\PineIR and the closest structure-elucidation methods. ``Extra input'' is the information that each model needs in addition to the spectrum, and ``Conformers'' states whether conformer ensembles are part of the training data. UV denotes ultraviolet spectra. MSD is a multimodal spectroscopic dataset, and QM9S and QMe14S are datasets of calculated spectra of small organic molecules.}
\label{tab:comparison}
\footnotesize
\renewcommand{\arraystretch}{1.25}
\setlength{\tabcolsep}{5pt}
\newcolumntype{Y}{>{\raggedright\arraybackslash\hyphenpenalty=10000}X}
\begin{tabularx}{\textwidth}{@{}l>{\raggedright\arraybackslash}p{2.6cm}>{\raggedright\arraybackslash}p{2.0cm}l>{\raggedright\arraybackslash}p{2.6cm}Yc@{}}
\toprule
\textbf{Model} & \textbf{Spectra} & \textbf{Extra input} & \textbf{Output} & \textbf{Paradigm} & \textbf{Spectral dataset} & \textbf{Conformers} \\
\midrule
SECS~\cite{mirzaEndtoendMultimodalStructure2026} & IR, \textsuperscript{1}H/\textsuperscript{13}C NMR, heteronuclear correlation & formula & ranked SMILES & retrieval + genetic search & MSD (794k)~\cite{NEURIPS2024_e38e60b3} & No \\
DiffSpectra~\cite{wangDiffSpectraMolecularStructure2025v2} & IR, Raman, UV & atom count & 2D + 3D & diffusion & QM9S (134k)~\cite{zouDeepLearningModel2023} & No \\
MAST~\cite{jiaMASTMotifAugmentedDiffusion2026} & IR, Raman, UV & atom count & 2D + 3D & diffusion + tree search & QM9S (134k)~\cite{zouDeepLearningModel2023} & No \\
IR-GeoDiff~\cite{wuLatentDiffusionbased3D2026} & IR & formula & 3D & latent diffusion & QM9S (134k)~\cite{zouDeepLearningModel2023} + \QMeS (50k)~\cite{yuanQMe14SComprehensiveEfficient} & No \\
\addlinespace[2pt]
\textbf{\PineIR} & IR & \textbf{none} & 3D & autoregressive latent diffusion & \PineDB (4.6M) & \textbf{Yes} \\
\bottomrule
\end{tabularx}
\end{table*}

In this work we show that 3D molecular structures can be generated from the IR spectrum alone, and that the generated geometries respond to the conformation encoded in the spectrum. The key is the simple observation that the IR spectrum of a molecule can be written exactly as a sum of atomic contributions. Our model, \PineIR, builds a molecule one atom at a time, and the contribution of each new atom to the spectrum is known. The number of atoms, the composition and the geometry are therefore determined during generation, and are not given as input. To train it, we computed the IR spectra of 4.64~M structures of 3.09~M molecules with the machine-learned interatomic potential SO3LR~\cite{kabyldaMolecularSimulationsPretrained2025}. To our knowledge, this is the largest dataset of calculated IR spectra, and the first that contains conformer ensembles. Note that all results below are obtained on computed spectra. The application to experimental spectra is discussed at the end of the paper.

\section{Results and discussion}
\label{sec:results}

\subsection{The IR spectrum as a sum of atomic contributions}
\label{sec:pir}

A harmonic calculation gives, for every normal mode $k$, a frequency $\nu_k$, an IR intensity $I_k$ and the displacement $\Delta\mathbf{r}_{ik}$ of every atom $i$. We distribute the intensity of each mode over the atoms according to their mechanical participation $w_{ik}$, in the spirit of earlier decompositions of normal modes~\cite{boatzDecompositionNormalcoordinateVibrational1989, jamrozVibrationalEnergyDistribution2013}:
\begin{align}
w_{ik} &= \frac{m_i \lVert \Delta\mathbf{r}_{ik} \rVert^2}{\sum_j m_j \lVert \Delta\mathbf{r}_{jk} \rVert^2}, \\
\mathrm{pIR}_i(\nu) &= \sum_k I_k\, w_{ik}\, L(\nu - \nu_k), \\
S(\nu) &= \sum_k I_k\, L(\nu - \nu_k) = \sum_{i=1}^{N} \mathrm{pIR}_i(\nu) ,
\end{align}
with $m_i$ the atomic mass, $N$ the number of atoms and $L$ a Lorentzian of 15\wn full width at half maximum on 3500 bins between 500 and 4000\wn. We call $\mathrm{pIR}_i$ the participation IR spectrum of atom $i$. Since $\sum_i w_{ik} = 1$, the molecular spectrum $S$ is exactly the sum of the atomic \pIR{}s. All spectra are normalised to unit mean intensity.

This decomposition plays the same role for IR spectra as the atomic energies of an MLIP play for the total energy~\cite{behlerGeneralizedNeuralNetworkRepresentation2007}. There is, however, an important difference. A \pIR is not a local property, since frequencies, intensities and weights all depend on the vibrations of the whole molecule. Predicting it from a finite atomic environment is therefore an approximation, and we test it in Section~\ref{sec:results_forward}.

\subsection{Generating a molecule atom by atom}
\label{sec:model}

\begin{figure*}[htb!]
    \centering
    \includegraphics[width=\textwidth]{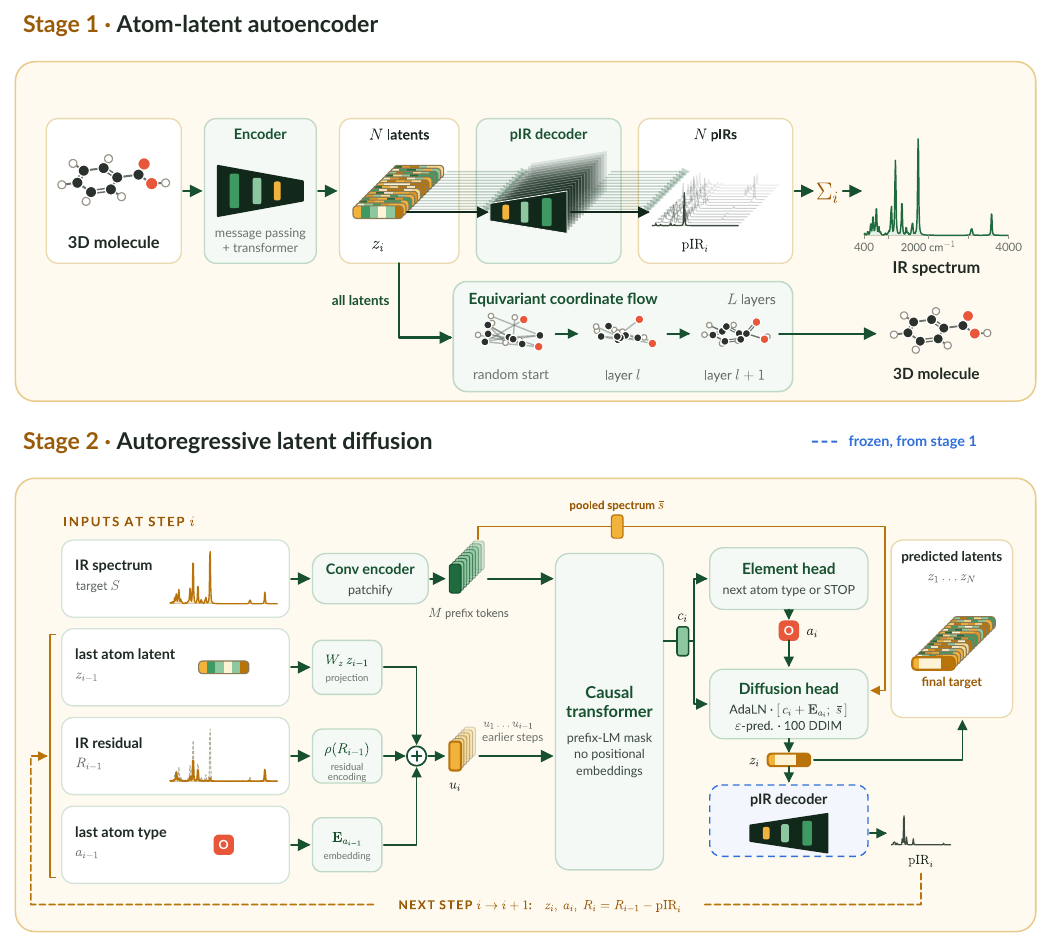}
    \caption{Architecture of \PineIR. Stage 1 encodes a molecule into one latent per atom and decodes its \pIR{}s and 3D structure. Stage 2 generates the latents from the spectrum, one atom at a time. Modules frozen after stage 1 are drawn in dashed blue.}
    \label{fig:pineir_steps}
\end{figure*}

\PineIR has two stages, trained one after the other (Figure~\ref{fig:pineir_steps}). The first is an autoencoder that maps every atom of a 3D structure to a latent vector. From the set of latents a structure decoder reconstructs the elements and the geometry, and a spectral decoder predicts the \pIR of each atom from its own latent. Each latent therefore describes both an atomic environment and its contribution to the spectrum.

\begin{figure*}[htb!]
    \centering
    \includegraphics[width=0.72\textwidth]{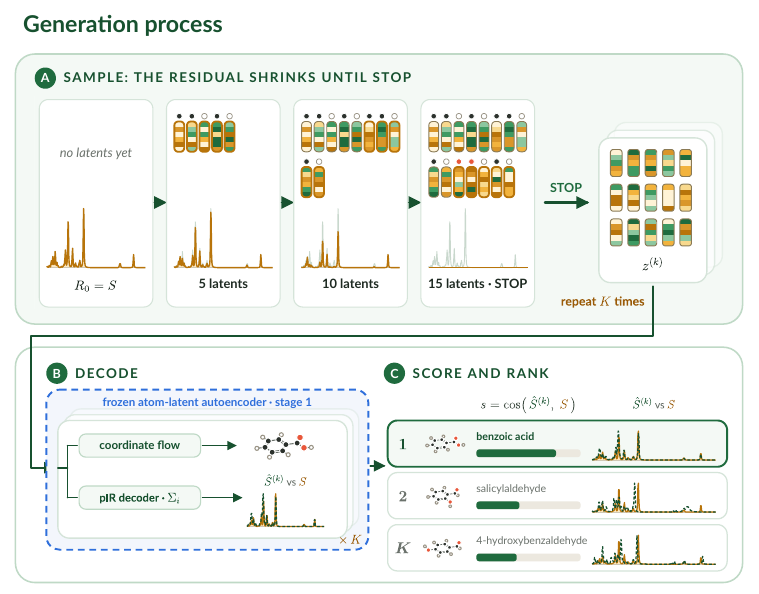}
    \caption{Generation of a molecule with \PineIR. The \pIR of each new latent is subtracted from the residual spectrum until the model generates \texttt{STOP}. The latents are then decoded into a 3D molecule, and the candidates are ranked by spectral cosine.}
    \label{fig:pineir_generation}
\end{figure*}

The second stage freezes the autoencoder and generates the latents from the spectrum, using the autoregressive latent diffusion approach developed for the generation of 3D molecules in KRONOS~\cite{ottomano2026autoregressive}. At each step a causal transformer chooses an element or a stop token, and a diffusion head samples the latent of the new atom. The \pIR of every generated atom is decoded and subtracted from the input, and the residual
\begin{equation}
R_i = S - \sum_{j \le i} \widehat{\mathrm{pIR}}_j
\end{equation}
is given to all later steps (Figure~\ref{fig:pineir_generation}). At each step the model therefore has access to the part of the spectrum that is not yet accounted for by the atoms already generated. The number of atoms and the elements are not given as input. When the model generates the stop token, the structure decoder converts the latents into a 3D molecule. The architecture and the training procedure are described in Supporting Information Section~{S6}.

We draw several candidates per spectrum and rank them by the cosine between the input spectrum and the sum of their decoded \pIR{}s, which we call the spectral cosine. Chemically stable candidates are ranked before unstable ones. This self-consistency criterion was proposed for IR-GeoDiff~\cite{wuLatentDiffusionbased3D2026}. In \PineIR it requires no additional model, since the generator already predicts the spectrum of every candidate.

\subsection{A database of 4.6 million IR spectra}
\label{sec:pinedbir}

\begin{figure*}[htb!]
    \centering
    \includegraphics[width=\textwidth]{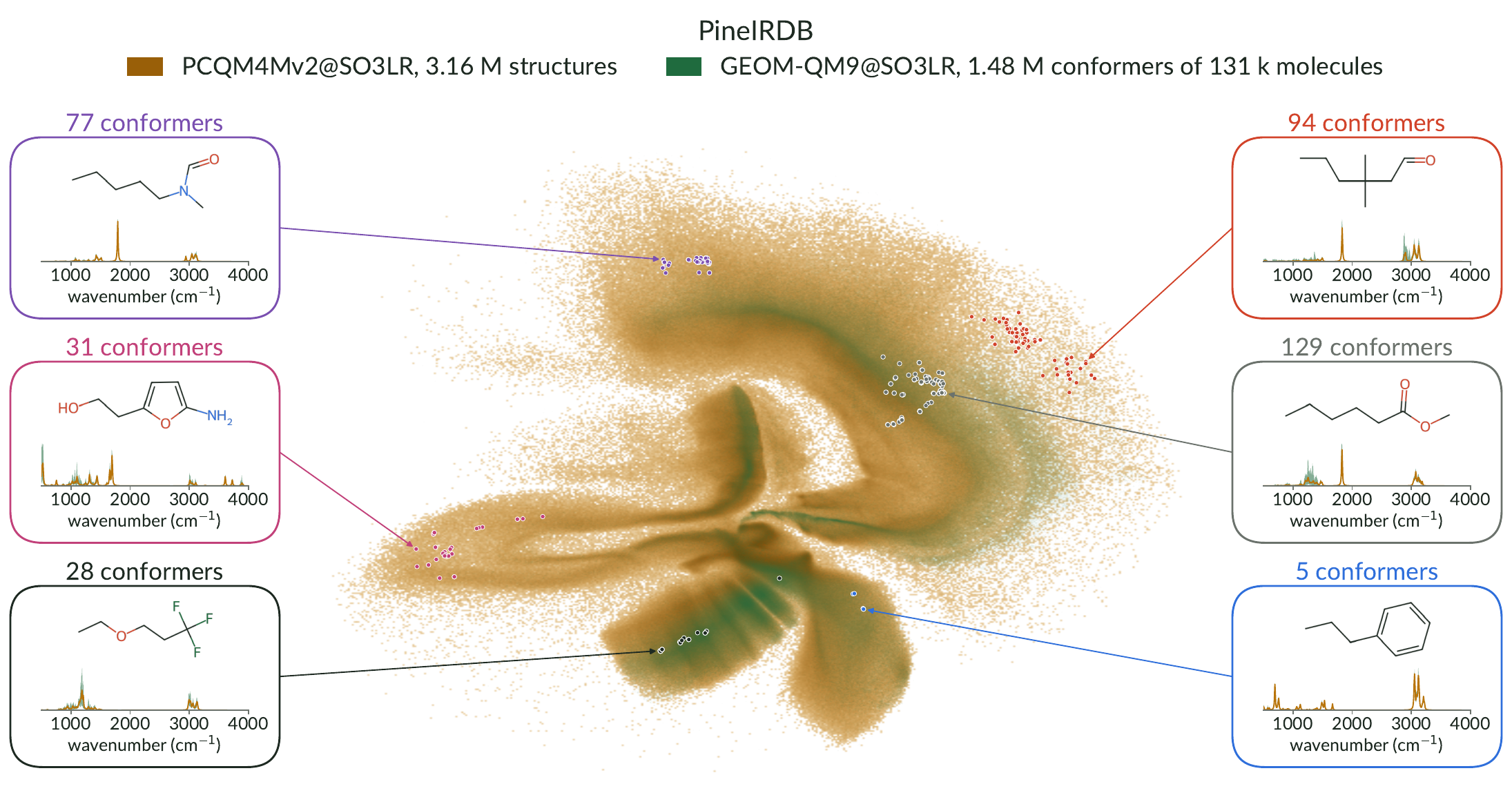}
    \caption{Two-dimensional map of the spectra in \PineDB. A separate variational autoencoder, used only for this map, compresses every spectrum into two numbers. The reconstructed spectra have a mean cosine of 0.78 with the original ones for 126{,}200 validation spectra, so the map is only qualitative. \PCQM (3.16~M structures) is shown in amber and \GEOM (1.48~M conformers of 131k molecules) in green, with a shading that grows with the logarithm of the density. Each card shows a molecule together with the mean (amber line) and the range (green band) of the spectra of its conformers.}
    \label{fig:latent_map}
\end{figure*}

To infer the composition, the model has to be trained on many different compositions, and to distinguish conformers it has to be trained on many conformers. We therefore computed harmonic IR spectra with SO3LR~\cite{kabyldaMolecularSimulationsPretrained2025} for structures from two databases (Table~\ref{tab:pinedbir}). \PCQM, a set of molecules from the PubChemQC project~\cite{nakataPubChemQCProjectLargeScale2017} released for the OGB large-scale challenge~\cite{huOGBLSCLargeScaleChallenge2021}, provides almost 3~M distinct molecules. \GEOM, the part of the GEOM dataset that contains the conformer ensembles of QM9 molecules~\cite{axelrodGEOMEnergyannotatedMolecular2022}, provides conformer ensembles, with 11.3 conformers per molecule on average and up to 1033. Together they form \PineDB. SO3LR was chosen after a benchmark against experimental gas-phase spectra (Supporting Information Section~{S3}), and it restricts the data to the eight elements H, C, N, O, F, P, S and Cl. Figure~\ref{fig:latent_map} shows a two-dimensional projection of the resulting spectra. The \PCQM spectra cover a broad region of this projection, while the spectra of the conformers of a single \GEOM molecule can be found in distinct regions. Finally, we use the B3LYP spectra of \QMeS, a dataset of calculated spectra of small organic molecules~\cite{yuanQMe14SComprehensiveEfficient} to test the transfer to a more accurate level of theory.

\begin{table}[htb!]
\caption{\PineDB and \QMeSDFT after cleaning, restricted to the eight SO3LR elements. Struct.\ is the number of structures, Conf./mol.\ the number of conformers per molecule, and Atoms the mean number of atoms per structure.}
\label{tab:pinedbir}
\footnotesize
\setlength{\tabcolsep}{3pt}
\begin{tabular}{llrrrr}
\toprule
Database & Spectra & Struct. & Molecules & Conf./mol. & Atoms \\
\midrule
\PCQM  & SO3LR      & 3.16~M & 2.97~M & 1.1  & 29.7 \\
\GEOM  & SO3LR      & 1.48~M & 131k   & 11.3 & 20.5 \\
\QMeS & B3LYP & 171k   & 168k   & 1.0  & 16.8 \\
\bottomrule
\end{tabular}
\end{table}

\begin{table*}[tb]
\caption{Correct SMILES (\%) on the external and in-distribution test sets, with 50 candidates per spectrum, grouped by the information each model receives besides the spectrum. All models are trained on the same \PineDB SO3LR split. ``First'' is the fraction of spectra for which the first candidate is correct, and ``50'' the fraction for which a correct candidate is found within 50 attempts. Chemically stable candidates are placed first, and no spectral ranking is used. The \PineIR rows with extra input are the same checkpoint sampled with that input imposed. Tables~{S4} and~{S5} give the matched RMSDs. ``Stable'' is the fraction of chemically stable candidates on the external set. Best value of each group in bold.}
\label{tab:results_qme14s_so3lr}
\label{tab:results_pineirdb_test}%
\setlength{\tabcolsep}{9pt}
\begin{tabular}{llrrrrr}
\toprule
 & & \multicolumn{2}{c}{External} & \multicolumn{2}{c}{In-distribution} & \\
\cmidrule(lr){3-4}\cmidrule(lr){5-6}
Extra input & Model & First $\uparrow$ & 50 $\uparrow$ & First $\uparrow$ & 50 $\uparrow$ & Stable \\
\midrule
none        & \PineIR     & 31 & 57 & 41 & 79 & 72 \\
\addlinespace
atom count  & \PineIR     & \textbf{37} & 58 & \textbf{51} & \textbf{79} & 50 \\
            & DiffSpectra & 22 & \textbf{63} & 22 & 75 & 85 \\
\addlinespace
formula     & \PineIR     & \textbf{45} & 72 & \textbf{50} & \textbf{80} & 57 \\
            & IR-GeoDiff  & 25 & \textbf{73} & 11 & 61 & 92 \\
\bottomrule
\end{tabular}
\end{table*}

\subsection{Structures without formula or atom count}
\label{sec:results_recovery}

We test \PineIR on two sets of molecules held out of training (see Methods), an \emph{external} set of 974 SO3LR spectra of 919 molecules from \QMeS, a separately curated database, and an \emph{in-distribution} set of 1000 molecules drawn at random from \PineDB. Every model generates 50 candidates per spectrum, and a candidate counts as correct when its canonical SMILES, without stereochemistry, matches the reference. The baselines, DiffSpectra~\cite{wangDiffSpectraMolecularStructure2025v2} and IR-GeoDiff~\cite{wuLatentDiffusionbased3D2026}, were retrained on the same \PineDB data.

Table~\ref{tab:results_qme14s_so3lr} gives the results. Using only the spectrum, \PineIR generates the correct molecule at the first attempt for 31\% of the external spectra and within 50 attempts for 57\%. For the in-distribution set these values increase to 41\% and 79\%.

The baselines cannot be used without extra input. For a fair comparison, we therefore also ran the same \PineIR model with the extra input imposed. The stop token is forbidden until the given atom count is reached, or elements are allowed only until the formula is used up. No retraining is involved. When the atom count is given, \PineIR generates the correct molecule at the first attempt for 37\% of the external spectra, against 22\% for DiffSpectra. When the formula is given, this value is 45\%, against 23\% for IR-GeoDiff. Within 50 attempts the differences are smaller. DiffSpectra performs better than \PineIR (63\% against 58\%), and IR-GeoDiff is level with it (73\% against 72\%). These are the only cases in the table in which a baseline matches or outperforms \PineIR. For the in-distribution set the differences are larger, and IR-GeoDiff, even with the formula, generates the correct molecule at the first attempt for only 11\% of the molecules.

Two further points are worth noting. First, for the in-distribution set the extra input does not improve the results of \PineIR within 50 attempts (79--80\% in all three cases), which shows that the model determines the composition correctly without this information. Second, imposing an extra input lowers the fraction of chemically stable candidates from 72\% to 50--57\% on the external set, because generation is forced to continue after the point where the model would stop. DiffSpectra and IR-GeoDiff produce more stable structures (85\% and 92\%), but chemical stability is only a necessary condition for a correct structure, and not a measure of the quality of the elucidation. When the molecule is correct, the geometry is close to the reference, with a root-mean-square deviation (RMSD) of 0.58--0.59~\AA\ for \PineIR on the external set, 0.59~\AA\ for DiffSpectra and 0.72~\AA\ for IR-GeoDiff.

\subsection{Ranking by the predicted spectrum}
\label{sec:results_ranking}

\begin{figure*}[p]
    \centering
    \includegraphics[width=\textwidth,height=0.84\textheight,keepaspectratio]{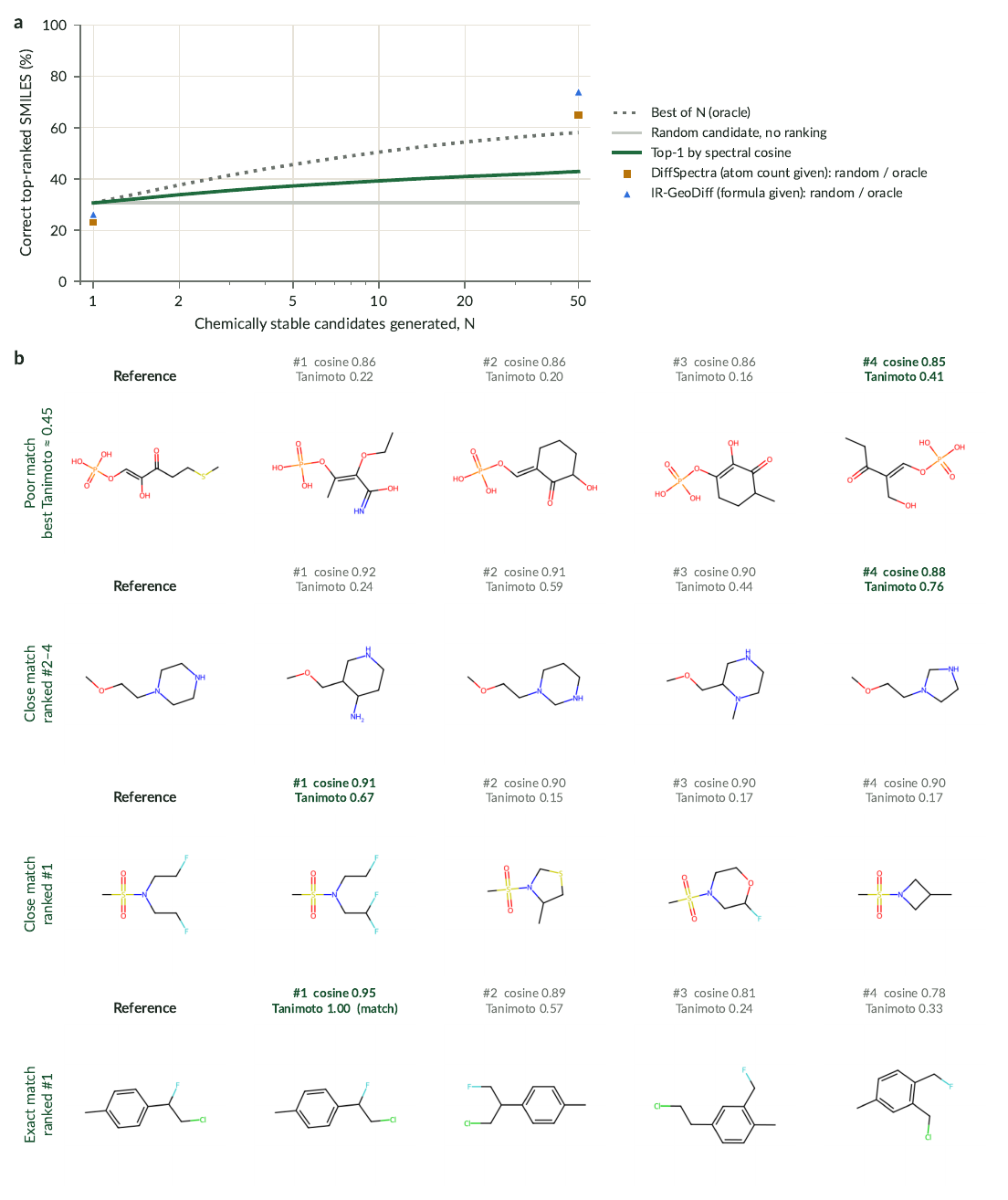}
    \caption{Ranking by spectral cosine on the external test set, against the number $N$ of chemically stable candidates per spectrum. Panel (a) shows the fraction of spectra for which the best-ranked candidate has the correct SMILES, compared with a randomly chosen candidate and with the best of the $N$ candidates. The curves are expected values over random subsets of the 50 stable candidates per spectrum, collected with at most 500 attempts. Markers give the random and best-of-50 values of DiffSpectra (given the atom count) and IR-GeoDiff (given the molecular formula). Panel (b) shows four test molecules and their four best-ranked distinct candidates, with the candidate closest to the reference in green.}
    \label{fig:realistic_generation}
\end{figure*}

A list of 50 candidates is only useful if the correct structure can be identified within it. Figure~\ref{fig:realistic_generation}a shows the effect of ranking by spectral cosine for the external set, as the pool of chemically stable candidates grows. A random candidate is correct for 31\% of the spectra. The best-ranked of ten candidates is correct for 39\% of the spectra, and the best-ranked of fifty candidates for 43\%. Since a correct candidate is present among the fifty for 58\% of the spectra, the remaining error is due to the ranking, and not to the generation.

When the best-ranked candidate is wrong, it is often chemically similar to the reference. For lists of ten candidates, the first-ranked candidate has a mean Morgan-fingerprint Tanimoto similarity of 0.53 to the reference, while the most similar of the ten has a similarity of 0.64 (Figure~{S7}). Figure~\ref{fig:realistic_generation}b shows typical cases, in which the candidates reproduce several functional groups of the reference but connect them differently. Randomly selected examples are in Figures~{S8} and~{S9}.

\subsection{Generated geometries reproduce the conformer and the diastereomer}
\label{sec:results_conformers}

A correct SMILES does not show that a model uses the geometric information contained in the spectrum. To test this, we use 291 held-out molecules with several SO3LR conformers (9085 spectra, the \emph{conformer} test set). For every correct candidate we compare its RMSD to the conformer that produced the input spectrum with its RMSD to other conformers of the same molecule. For a model that ignores the 3D geometry, the generated structure is closer to the input conformer in 50\% of the cases.

\begin{figure*}[tp]
    \centering
    \includegraphics[width=\textwidth,height=0.4\textheight,keepaspectratio]{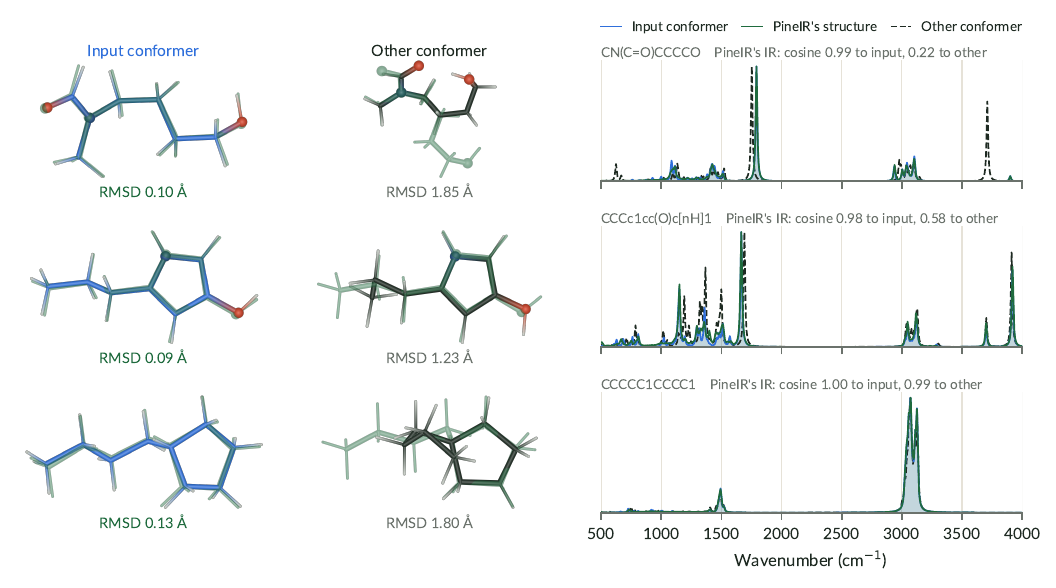}
    \caption{Three molecules of the conformer test set. The best-ranked correct \PineIR candidate (green) is overlaid on the input conformer (blue) and on another conformer (dark), and their IR spectra are shown on the right. RMSDs are over the whole molecule, and the overlays align a shared fragment.}
    \label{fig:conformers}

    \vspace{1.5em}
    \includegraphics[width=\textwidth,height=0.4\textheight,keepaspectratio]{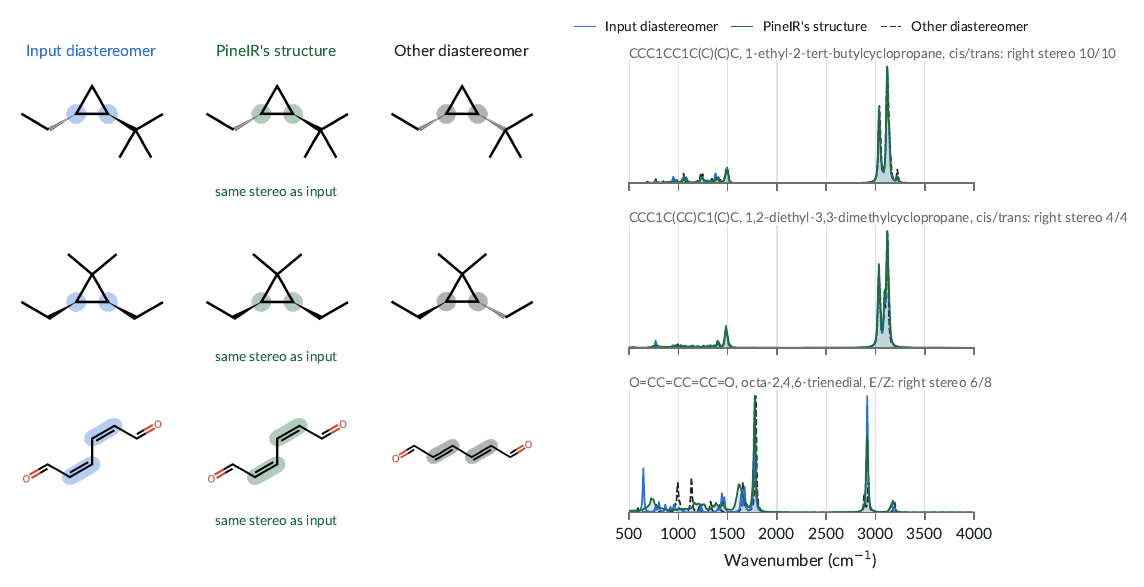}
    \caption{Diastereomers of 1-ethyl-2-\textit{tert}-butylcyclopropane and 1,2-diethyl-3,3-dimethylcyclopropane (cis/trans, conformer test set) and octa-2,4,6-trienedial (E/Z, external test set). For each molecule we show the diastereomer used as input (blue), the best-ranked correct candidate (green) and the other diastereomer (dark), together with their IR spectra. The fraction is the percentage of correct \PineIR candidates with the stereochemistry of the input.}
    \label{fig:stereo}
\end{figure*}

\PineIR's candidates are closer to the input conformer in 78\% of the comparisons, against 68\% for DiffSpectra and 58\% for IR-GeoDiff, only slightly above a random choice (Table~\ref{tab:conformer_swap}). The mean RMSD to the input conformer is 0.58~\AA\ for \PineIR, against 0.99~\AA\ to the other conformers. Figure~\ref{fig:conformers} shows three examples.

\begin{table}[htb!]
\caption{Conformer sensitivity on the conformer test set. Each spectrum is paired with up to 10 other conformers of the same molecule (RMSD $>0.1$~\AA). ``Closer'' is the fraction of comparisons in which a correct candidate is closer to the input conformer (50\% for a random choice). RMSDs in \AA. Each model is evaluated on the molecules for which it generates the correct SMILES (267, 265 and 257 of 291). 95\% bootstrap intervals in brackets.}
\label{tab:conformer_swap}
\footnotesize
\setlength{\tabcolsep}{3pt}
\begin{tabular}{lccc}
\toprule
Model & RMSD input & RMSD other & Closer (\%) \\
\midrule
DiffSpectra & 0.73 [0.69, 0.78] & 0.98 [0.93, 1.01] & 68 [66, 70] \\
IR-GeoDiff  & 0.96 [0.92, 0.99] & 1.04 [1.00, 1.08] & 58 [56, 59] \\
\textbf{\PineIR} & 0.58 [0.54, 0.61] & 0.99 [0.95, 1.03] & 78 [76, 81] \\
\bottomrule
\end{tabular}
\end{table}

Diastereomers provide a more stringent test. Unlike enantiomers, they have different IR spectra, so the spectrum determines whether a ring substituent is cis or trans, or whether a double bond is E or Z. On the 161 molecules of the conformer set with at least two diastereomers, \PineIR returns the diastereomer of the input in 83\% of its correct candidates, against 72\% for DiffSpectra, 60\% for IR-GeoDiff and 37\% for a random choice. The same trend is found for the external set (84\%, 74\% and 64\% for 142 molecules). A high recovery rate could, however, also be obtained by a model that always returns the more common diastereomer. To exclude this possibility, we use the fifteen molecules for which both diastereomers are present in the test set. For these, a model that ignores the spectrum returns the correct diastereomer in half of the cases, while \PineIR does so in 72\% [63, 83] of the cases, DiffSpectra in 66\% [58, 73] and IR-GeoDiff in 52\% [42, 64]. Figure~\ref{fig:stereo} shows three such pairs.

\subsection{A fast forward model with atomic resolution}
\label{sec:results_forward}

The first stage of \PineIR can also be used on its own as a forward model. From a 3D structure it predicts the \pIR of every atom, and their sum is the molecular spectrum. Several neural networks predict IR spectra from the molecular structure, most of them from the 2D molecular graph~\cite{mcgillPredictingInfraredSpectra2021, Saquer2024AttentiveFPIR, ShabanTameh2024CMPNNIR, stienstraGraphormerIRGraphTransformers2024, naDeepLearningGenerating2024, Liu2026MolSpectra, Stienstra2025GraphormerIRIS}. NNMol-IR uses 3D coordinates and sums atomic contributions, but these contributions are not part of the training targets~\cite{AbdulAl2024NNMolIR}. In \PineIR they are, so the predictions can be compared with the reference for each atom.

On the external set the predicted spectra have a median cosine of 0.86 (mean 0.73) and a mean spectral information similarity (SIS)~\cite{mcgillPredictingInfraredSpectra2021} of 0.66 with the SO3LR spectra. A prediction takes 8.4~ms on a graphics processing unit (GPU), against 2.35~s for the SO3LR harmonic calculation of the same geometry. The two timings were measured on different GPUs, so the factor of about 280 is only indicative.

\begin{figure}[htb!]
    \centering
    \includegraphics[width=\columnwidth]{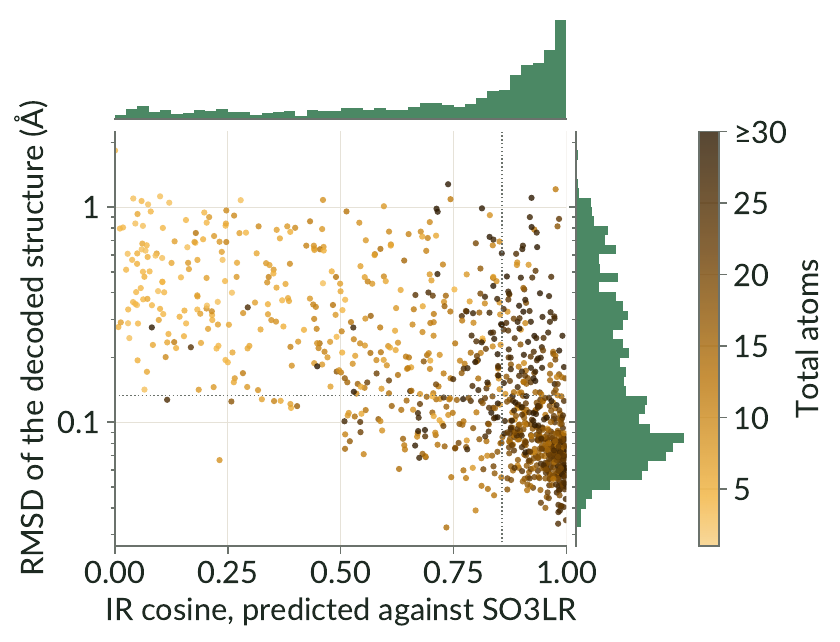}
    \caption{Spectral error (cosine between predicted and SO3LR spectra) against structural error (RMSD of the structure decoded from the same latents) for the 969 IR-active structures of the external set, coloured by atom count. The dotted lines mark the medians (cosine 0.86, RMSD 0.13~\AA).}
    \label{fig:forward_errors}
\end{figure}

\begin{figure}[htb!]
    \centering
    \includegraphics[width=\columnwidth]{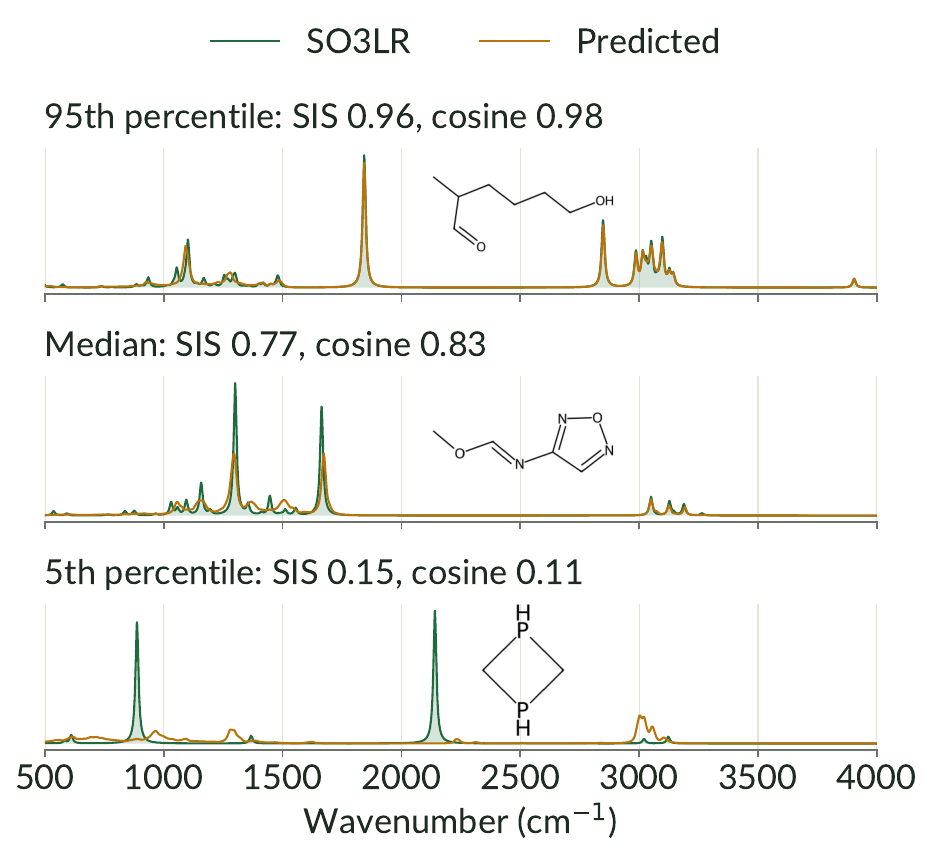}
    \caption{Predicted and SO3LR IR spectra at the 95th-percentile, median and 5th-percentile SIS of the external set. Spectra are area-normalised.}
    \label{fig:forward_examples}
\end{figure}

The largest errors are found for two classes of molecules (Figures~\ref{fig:forward_errors} and~\ref{fig:forward_examples}). The first are small molecules, which are rare in the training set. For molecules with up to 7 heavy atoms the median cosine is 0.46, against 0.95 at 8--9 heavy atoms. For these molecules the structure decoded from the same latents also has larger errors (median RMSD 0.31~\AA\ against 0.08~\AA). The second are molecules containing phosphorus, an element present in 1\% of the training molecules but in 17\% of the external set. At 8--9 heavy atoms, molecules with phosphorus have a median SIS of 0.39, against 0.88 for those without. In the worst cases the model does not reproduce the P--H stretching band near 2140\wn, and assigns its intensity to the C--H stretching bands instead. In fact, this band is more than 350 times weaker than the C--H maximum in the average training spectrum (Figure~{S4}).

\begin{figure}[htb!]
    \centering
    \includegraphics[width=\columnwidth]{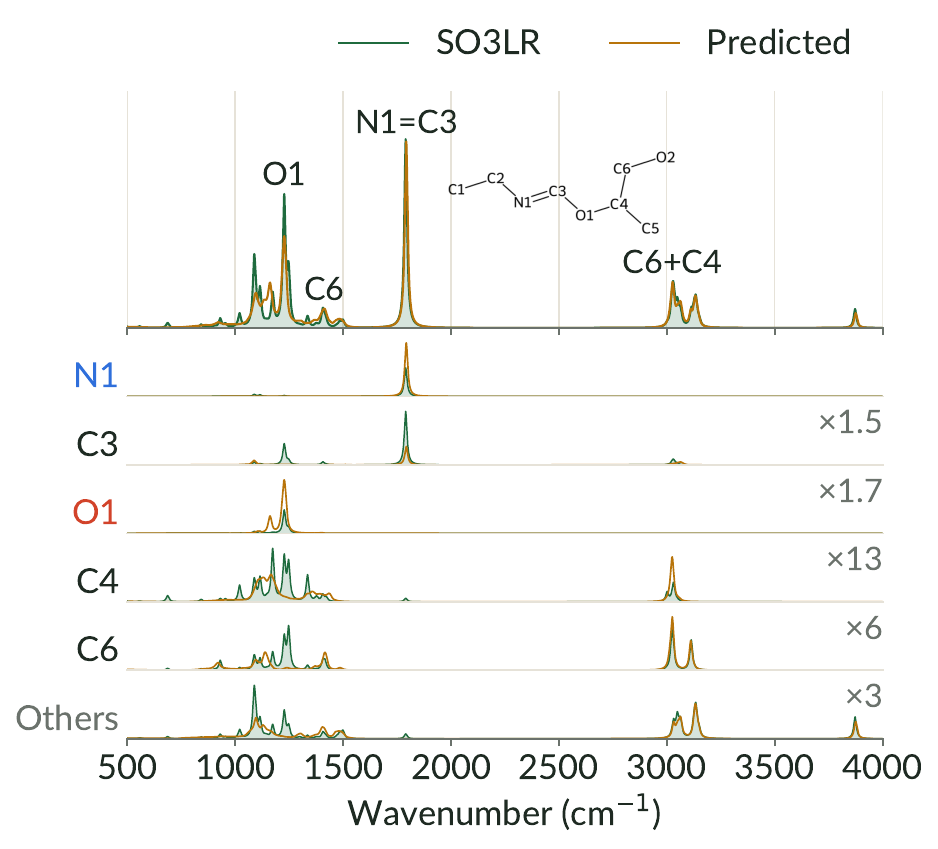}
    \caption{Predicted (amber) and SO3LR (green) \pIR{}s of a test molecule. Each heavy atom is shown together with its hydrogen atoms, and the sum of all contributions gives the molecular spectrum at the top. Each row is scaled to its maximum and magnified by the factor given on its right. Bands are labelled with the atoms that contribute most to them, and N1=C3 denotes the C=N stretching band.}
    \label{fig:forward_assignment}
\end{figure}

Figure~\ref{fig:forward_assignment} shows the atom-resolved prediction for one test molecule. For this molecule the predicted and reference \pIR{}s agree closely, so the model can be used to assign the bands of a spectrum to atoms and functional groups. Note that this assignment depends on our choice of partition of the intensity. The normal modes are a property of the whole molecule, and their partition into atomic contributions is not unique.

\subsection{Transfer to DFT spectra}
\label{sec:results_dft}

SO3LR is trained on data calculated with the hybrid functional PBE0~\cite{kabyldaMolecularSimulationsPretrained2025}, while B3LYP reproduces experimental frequencies better~\cite{baroneFullyAnharmonicIR2014}. The B3LYP spectra of \QMeSDFT therefore test whether \PineIR can be transferred to a more accurate level of theory. Without fine-tuning this is not the case. For the \QMeSDFT validation molecules it generates the correct molecule at the first attempt for 2\% of the spectra, against 49\% for the same molecules with SO3LR spectra. The differences between the two levels of theory are therefore large enough to make the model fail.

We then fine-tuned \PineIR on the \QMeSDFT training split (161k structures, about 30 times smaller than \PineDB) and tested it on its 1000 test molecules (Table~\ref{tab:results_dft}). The fine-tuned model generates the correct molecule at the first attempt for 32\% of the molecules, against 23\% for the same architecture trained on \QMeSDFT alone. The paired difference is $+9$ percentage points (95\% confidence interval [$+6$, $+12$]), and $+14$ [$+10$, $+18$] on the 527 molecules made only of H, C, N, O and F (HCNOF in Table~\ref{tab:results_dft}), where the fine-tuned model reaches 50\%. Pretraining on \PineDB is therefore useful even when the target level of theory is different.

Fine-tuning, however, has a drawback. After fine-tuning, \PineIR generates many identical candidates, with only 6.3 distinct structures among 50 attempts, against 14.9 for DiffSpectra. As a consequence, DiffSpectra, which performs similarly at the first attempt (34\%), performs much better within 50 attempts (77\% against 46\%). The published checkpoints of DiffSpectra and IR-GeoDiff were trained on the QM9S dataset, which likely contains part of this test set, so we report them only in Supporting Information Section~{S5}.

\begin{table}[htb!]
\caption{Correct SMILES (\%) on the \QMeSDFT test set (B3LYP spectra), 50 candidates per spectrum, scored as in Table~\ref{tab:results_qme14s_so3lr}. Models marked $\rightarrow$ are pretrained on \PineDB and fine-tuned on \QMeSDFT, and ``scratch'' is trained on \QMeSDFT only. DiffSpectra receives the atom count.}
\label{tab:results_dft}
\footnotesize
\setlength{\tabcolsep}{4pt}
\begin{tabular}{llrrrr}
\toprule
 & & \multicolumn{2}{c}{All (1000)} & \multicolumn{2}{c}{HCNOF (527)} \\
\cmidrule(lr){3-4}\cmidrule(lr){5-6}
Model & Training & First & 50 & First & 50 \\
\midrule
DiffSpectra & $\rightarrow$ & \textbf{34} & \textbf{77} & 43 & \textbf{90} \\
\PineIR     & scratch       & 23 & 34 & 36 & 51 \\
\PineIR     & $\rightarrow$ & 32 & 46 & \textbf{50} & 66 \\
\bottomrule
\end{tabular}
\end{table}

\section{Conclusions}
\label{sec:conclusion}

In conclusion, we presented \PineIR, a generative model that returns 3D molecular structures from an IR spectrum without the molecular formula or the atom count. It builds the molecule atom by atom, using the fact that the IR spectrum is a sum of atomic contributions. On computed spectra of held-out molecules, the best-ranked of fifty candidates is correct for 43\% of the spectra, and a correct candidate is among them for 58\%. When given the same extra input as DiffSpectra or IR-GeoDiff, \PineIR generates the correct molecule at the first attempt more often than both. The generated geometries reproduce the conformer and the diastereomer of the input spectrum, which shows that the IR spectrum carries information on the 3D geometry, and that \PineIR uses it. In addition, the first stage of the model predicts molecular and atom-resolved IR spectra in milliseconds.

The model has clear limitations. Its candidates are less diverse than those of DiffSpectra, especially after fine-tuning, and it is therefore outperformed within 50 attempts even in cases where it performs better at the first attempt. Small molecules and phosphorus compounds are underrepresented in training and are predicted worse. The training data contain only eight elements and molecules with up to 53 atoms, and correspond to isolated molecules in the harmonic approximation. Most importantly, all tests use calculated spectra. The failure of the SO3LR model for B3LYP spectra without fine-tuning shows that it cannot be applied directly to measured spectra. We believe that fine-tuning provides a solution to this problem. A set of 161k B3LYP spectra was sufficient to adapt the pretrained model to a new level of theory, and a curated set of experimental spectra of similar size should be sufficient to adapt it to experiment. The construction of such a set is the next step of this work. Finally, since the number of atoms is not fixed in advance, the method can in principle be extended to mixtures of molecules.

\section{Methods}
\label{sec:methods}

\paragraph{Data.} Structures from \PCQM~\cite{huOGBLSCLargeScaleChallenge2021} and \GEOM~\cite{axelrodGEOMEnergyannotatedMolecular2022} were optimised with SO3LR~\cite{kabyldaMolecularSimulationsPretrained2025}, followed by a harmonic analysis. We kept converged structures without imaginary frequencies, removed disconnected or chemically invalid graphs with the criteria of Nikitin \textit{et al.}~\cite{nikitinGEOMdrugsRevisitedMore2025}, and removed geometric duplicates. We use the same valence and connectivity criteria to decide whether a generated molecule is chemically stable. Supporting Information Sections~{S1} and~{S3} describe the cleaning procedure and the benchmark of the potential.

\paragraph{Test sets.} All conformers of a molecule are assigned to the same partition. The \emph{external} set is a 1000-molecule test split of \QMeS, chosen to cover size, elements and functional groups, and held out of \PineDB by canonical SMILES. After SO3LR re-optimisation 919 molecules (974 structures) remain. The \emph{in-distribution} set holds 1000 further molecules drawn at random from \PineDB (901 from \PCQM, 99 from \GEOM). The \emph{conformer} set holds the \PineDB conformers of the 510 \QMeS test molecules that \PineDB contains, and its 291 molecules with at least two conformers are used in Table~\ref{tab:conformer_swap}. The \emph{DFT} test set is the same \QMeS split with its original spectra, calculated with B3LYP and a TZVP basis set, and contains all 1000 molecules. Coverage diagnostics are in Figure~{S5}. The evaluation sets, the protocols and the full results tables are given in Supporting Information Section~{S2}.

\paragraph{Metrics.} A candidate is correct when it is chemically stable and its canonical SMILES, without stereochemistry, equals the reference. Unstable candidates count as failures. In the fixed-attempt benchmarks (Tables~\ref{tab:results_qme14s_so3lr} and~\ref{tab:results_dft}) stable candidates are placed first, in generation order, and no spectral ranking is used. The ranking analysis (Figure~\ref{fig:realistic_generation}) collects 50 stable candidates with at most 500 attempts. RMSDs use a graph mapping and a Kabsch alignment~\cite{kabschSolutionBestRotation1976} with reflections allowed. The stereochemistry is determined from the 3D coordinates, and mirror images are counted as one.

\paragraph{Baselines.} DiffSpectra~\cite{wangDiffSpectraMolecularStructure2025v2} receives the IR spectrum and the atom count, and IR-GeoDiff~\cite{wuLatentDiffusionbased3D2026} the IR spectrum and the formula. Both were retrained on the same \PineDB split. Checkpoints are listed in Supporting Information Section~{S7}.

\section*{Data and code availability}
Code and data will be released upon publication.

\section*{Author contributions}
G.B. and M.A.L.M. conceived the project. P.-P.D.B. designed and developed the initial machine-learning framework and supervised the later training and architecture experiments. G.B. built the datasets, developed the later stages of the model architecture, ran the tests and wrote the paper. S.B. and M.A.L.M. supervised the project and contributed to writing the paper.

\section*{Acknowledgements}
We are grateful to Thalis da Silva, Paulo Pires and Nicolas Tancogne-Dejean for helpful discussions. Parts of the calculations were performed on the high-performance computing cluster Elysium of the Ruhr University Bochum, subsidised by the Deutsche Forschungsgemeinschaft (DFG) (INST 213/1055-1).

\section*{Supporting Information}
The Supporting Information describes the preparation, cleaning, splitting and coverage of the datasets, the evaluation sets and protocols with the full results tables, the benchmark of the machine-learned potential, the architecture and training of both stages, and the checkpoints behind every result. It also contains randomly selected examples and the results of the published checkpoints of DiffSpectra and IR-GeoDiff.

\clearpage
\renewcommand{\bibsection}{\section*{References}}
\bibliographystyle{achemso}
\bibliography{library}

\end{document}

\newcommand{\AffOne}{Theoretical Chemistry, VU University Amsterdam, The Netherlands}
\newcommand{\AffTwo}{Research Center Future Energy Materials and Systems of the University Alliance Ruhr and Interdisciplinary Centre for Advanced Materials Simulation, Ruhr University Bochum, Universitätsstraße 150, D-44801 Bochum, Germany}

\renewcommand{\thefigure}{S\arabic{figure}}
\renewcommand{\thetable}{S\arabic{table}}
\renewcommand{\thepage}{S\arabic{page}}
\renewcommand{\theequation}{S\arabic{equation}}
\renewcommand{\thesection}{S\arabic{section}}

\title{Supporting Information for ``PineIR: 3D molecular structures from an infrared spectrum alone''}

\author{Giulio Benedini\,\orcidlink{0009-0000-0943-8744}}
\affiliation{\AffOne}
\affiliation{\AffTwo}

\author{Pierre-Paul De Breuck\,\orcidlink{0000-0002-3173-2058}}
\affiliation{\AffTwo}

\author{Miguel A. L. Marques\,\orcidlink{0000-0003-0170-8222}}
\affiliation{\AffTwo}

\author{Silvana Botti\,\orcidlink{0000-0002-4920-2370}}
\email{silvana.botti@rub.de}
\affiliation{\AffTwo}

\date{\today}

\maketitle

\begin{center}
    \textbf{Supporting Information}
\end{center}

Section~\ref{sec:datasets} describes the preparation, cleaning, splitting and coverage of the datasets. Section~\ref{sec:evaluation} defines the evaluation sets and protocols, and gives the full results tables. Section~\ref{sec:mlip_choice} presents the benchmark used to select the machine-learned potential, and Section~\ref{sec:examples} shows randomly selected structure-elucidation examples. Section~\ref{sec:published_checkpoints} reports the published checkpoints of DiffSpectra and IR-GeoDiff, Section~\ref{sec:architecture} the architecture and training objectives of both stages, and Section~\ref{sec:checkpoints} the checkpoints behind every result.

\listoffigures
\listoftables

\clearpage

\section{Dataset preparation and split diagnostics}
\label{sec:datasets}

To trace how common molecules retain their split across databases, Figure~\ref{fig:cleaning_workflow} summarises cleaning and partitioning.

\begin{figure}[htb!]
    \centering
    \begin{tikzpicture}[
    font=\footnotesize,
    node distance=4mm,
    >={Stealth[length=2mm]},
    flow/.style={->, draw=pineirInk, line width=0.6pt},
    reuse/.style={->, draw=pineirInk, line width=0.6pt, dashed},
    link/.style={<->, draw=pineirTest, line width=0.6pt, dashed},
    edgelabel/.style={font=\scriptsize, text=pineirMuted, fill=white, inner sep=1pt},
    box/.style={draw=pineirMuted, fill=pineirFill, rounded corners=2pt,
                 align=center, inner sep=4pt},
    db/.style={cylinder, shape border rotate=90, aspect=0.12, draw=#1,
               fill=#1!12, line width=0.8pt, align=center,
               minimum width=38mm, inner sep=3pt},
    test/.style={draw=pineirTest, fill=pineirTest!10, line width=0.8pt,
                 rounded corners=2pt, align=center, minimum width=47mm,
                 minimum height=13mm, inner sep=3pt},
    frame/.style={draw=pineirMuted, rounded corners=4pt, inner sep=7pt},
    framelabel/.style={font=\scriptsize\itshape, text=pineirMuted, anchor=north west},
]
\def\xl{-5.6}
\def\xm{0}
\def\xr{5.6}

\node[db=pineirNeutral] (Q) at (\xl, 0) {\textbf{\QMeS}\\186k molecules\\B3LYP/TZVP structures + IR};
\node[db=pineirNeutral] (G) at (\xm, 0) {\textbf{\GEOM}\\1.8\,M conformers\\of 133k molecules};
\node[db=pineirNeutral] (P) at (\xr, 0) {\textbf{\PCQM}\\3.5\,M molecules\\[0.5ex]};

\node[box, text width=148mm] (EL) at (0, -1.75) {Keep only H, C, N, O, F, P, S, Cl};
\foreach \s in {Q, G, P} \draw[flow] (\s.south) -- (\s.south |- EL.north);

\node[box, draw=pineirSOLR, fill=pineirSOLR!12, text width=95mm] (SO)
    at (2.8, -3.05) {\textbf{SO3LR} geometry optimisation + harmonic IR\\(\QMeS, \GEOM, \PCQM)};
\draw[flow] (SO.north |- EL.south) -- (SO.north);

\begin{scope}[every node/.style={box, text width=33mm, minimum height=12mm}]
    \node (F1) at (-5.55, -5.05) {1\,$\cdot$\,Imaginary frequencies\\drop if $\nu_\text{min} < 0$};
    \node (F2) at (-1.85, -5.05) {2\,$\cdot$\,Molecular graph\\RDKit +\\extended H\"uckel};
    \node (F3) at ( 1.85, -5.05) {3\,$\cdot$\,Valid graphs\\drop fragments and\\invalid valences};
    \node (F4) at ( 5.55, -5.05) {4\,$\cdot$\,Duplicates\\same SMILES and\\RMSD $< 0.05$\,\AA};
\end{scope}
\foreach \a/\b in {F1/F2, F2/F3, F3/F4} \draw[flow] (\a) -- (\b);
\begin{scope}[on background layer]
    \node[framelabel, anchor=south] (FL) at ($(F1.north -| -1.4,0) + (0,1mm)$) {Filters, applied to each database in order};
    \node[frame, fit=(F1)(F4)(FL), inner xsep=7pt, inner ysep=4pt] (FIL) {};
\end{scope}

\draw[flow] (Q |- EL.south) -- node[edgelabel, pos=0.45] {\QMeS (B3LYP/TZVP)} (Q |- FIL.north);
\draw[flow] (SO.south) -- (SO.south |- FIL.north);

\node[db=pineirQMeSDFT] (QO) at (\xl, -7.55) {\textbf{\QMeSDFT}\\171k structures};
\node[db=pineirSOLR]    (QS) at (\xm, -7.55) {\textbf{\QMeSSO}\\164k structures};
\node[db=pineirSOLR]    (PI) at (\xr, -7.55) {\textbf{\PineDB}\\4.64\,M structures\\of 3.09\,M molecules};
\foreach \d in {QO, QS, PI} \draw[flow] (\d.north |- FIL.south) -- (\d.north);

\node[box, text width=148mm] (SP) at (0, -9.6)
    {\textbf{Split \QMeSDFT by canonical SMILES}: test 1{,}000 molecules, validation 5\%, training the rest\\
     \QMeSSO and \PineDB reuse this partition for every molecule they share with \QMeSDFT;\\
     Remove shared \QMeSDFT test molecules from \PineDB, then randomly select 1{,}000 more for test};
\draw[flow]  (QO.south) -- (QO.south |- SP.north);
\draw[reuse] (QS.south) -- node[edgelabel] {reuse partition} (QS.south |- SP.north);
\draw[reuse] (PI.south) -- node[edgelabel] {reuse partition} (PI.south |- SP.north);

\node[test] (TD) at (\xl, -11.75) {\textbf{\QMeSDFT-test}\\DFT (B3LYP/TZVP)\\1{,}000 molecules};
\node[test] (TS) at (\xm, -11.75) {\textbf{\QMeSSO-test}\\919 molecules; conformer subset:\\510 shared molecules, 9{,}304 structures};
\node[test] (TP) at (\xr, -11.75) {\textbf{\PineDB-test}\\SO3LR, randomly selected\\1{,}000 molecules};
\foreach \t in {TD, TS, TP} \draw[flow] (\t.north |- SP.south) -- (\t.north);
\draw[link] (TD) -- (TS);
\begin{scope}[on background layer]
    \node[frame, draw=pineirTest, fit=(TD)(TP), inner ysep=6pt] (TEST) {};
\end{scope}
\node[framelabel, text=pineirTest, anchor=north] at (TEST.south) {QMe14S test molecules are shared across levels of theory; the conformer subset comes from PineIRDB};
\end{tikzpicture}
    \caption[Cleaning and splitting workflow.]{Cleaning and splitting workflow. \QMeS is used twice, with its B3LYP/TZVP spectra (\QMeSDFT) and recomputed with SO3LR (\QMeSSO). The split is drawn on \QMeSDFT, and \QMeSSO and \PineDB reuse its partition for every molecule they share with it. The \QMeS test molecules can therefore be evaluated at either level of theory; \PineDB holds 510 of the shared test molecules, the \QMeS conformer set; the 356 among them from \GEOM supply its conformers. After removing the shared \QMeS test molecules, 1000 further \PineDB molecules are randomly selected for \PineDB-test.}
    \label{fig:cleaning_workflow}
\end{figure}

\subsection{Available IR data}
\label{sec:available_ir_data}

Public IR data, summarised in Table~\ref{tab:public_ir_datasets}, fall into two groups. Experimental collections (the NIST Chemistry WebBook~\cite{nistChemistryWebBook}, $\sim$16k spectra; the Graphormer-IR compilation~\cite{stienstraGraphormerIRGraphTransformers2024}, 54k spectra in five measurement media) are the ground truth, but they are small, heterogeneous in phase and technique, and mostly not freely redistributable. Computed collections are larger and internally consistent, but each trades away one of the properties we need. The QM9-derived sets, QM9S~\cite{zouDeepLearningModel2023} and SQuIRL~\cite{krishnadasSpectralQuantumChemistry2026}, provide harmonic DFT spectra for $\sim$134k molecules of at most nine heavy atoms drawn from five elements. \QMeS~\cite{yuanQMe14SComprehensiveEfficient} is the widest DFT set available, with 186k molecules at the B3LYP/TZVP level spanning 14 elements and up to 32 heavy atoms, and is the reference used by recent IR-to-structure models~\cite{wangDiffSpectraMolecularStructure2025v2, wuLatentDiffusionbased3D2026}. Larger sets give up accuracy: MSD~\cite{NEURIPS2024_e38e60b3} reaches 794k molecules only by using a classical force field, which falls well short of quantum-chemical accuracy for IR spectra~\cite{henschelTheoreticalInfraredSpectra2020}, and the 495k IR/Raman pairs of Mitsuishi \textit{et al.}~\cite{mitsuishiDeepLearningAssistedNovo2025} remain within organic chemistry of modest size. Specialised sets such as the in-silico library of molecular ions for metabolite identification~\cite{houthuijsSilicoInfraredSpectral2023} cover a different problem entirely.

\begin{table}[htb!]
\caption{Public IR datasets, computed and experimental, with the role each plays in this work.}
\label{tab:public_ir_datasets}
\begin{tabular}{p{2.7cm}p{1.9cm}p{2.1cm}p{2.9cm}p{3.0cm}p{3.0cm}}
\toprule
Dataset & Kind & Size & Level / source & Chemical space & Used in \\
\midrule
NIST WebBook~\cite{nistChemistryWebBook} & experimental & $\sim$16k spectra & gas-phase and condensed-phase measurements & broad organic, many elements & SMEN~\cite{kanakalaSpectraStructureContrastive2024} (27 spectra) \\
\addlinespace
Graphormer-IR set~\cite{stienstraGraphormerIRGraphTransformers2024} & experimental & 54k spectra & five measurement media & H, C, N, O, F, Si, S, P, Cl, Br, I & forward IR prediction \\
\addlinespace
VIBFREQ1295~\cite{zapatatrujilloVIBFREQ1295NewDatabase2022} & computed + experimental & 141 molecules, 1295 modes & CCSD(T)(F12*)/cc-pVDZ-F12 harmonic + experimental fundamentals & $\le 6$ heavy atoms & our MLIP frequency benchmark \\
\addlinespace
QM9S~\cite{zouDeepLearningModel2023} & computed & 134k molecules & harmonic DFT & QM9 (H, C, N, O, F; $\le 9$ heavy atoms) & DiffSpectra~\cite{wangDiffSpectraMolecularStructure2025v2}, IR-GeoDiff~\cite{wuLatentDiffusionbased3D2026} \\
\addlinespace
SQuIRL~\cite{krishnadasSpectralQuantumChemistry2026} & computed & 133{,}885 molecules & $\omega$B97X-D/aug-cc-pVTZ harmonic & QM9 & --- \\
\addlinespace
\QMeS~\cite{yuanQMe14SComprehensiveEfficient} & computed & 186k molecules & B3LYP/TZVP harmonic & 14 elements, 47 functional groups, $\le 32$ heavy atoms & IR-GeoDiff~\cite{wuLatentDiffusionbased3D2026}; our DFT reference \\
\addlinespace
Mitsuishi \textit{et al.}~\cite{mitsuishiDeepLearningAssistedNovo2025} & computed & 495k IR + Raman pairs & DFT & organic & formula-free SMILES generation \\
\addlinespace
MSD~\cite{NEURIPS2024_e38e60b3} & computed & 794k molecules & classical force-field MD & organic molecules & SECS~\cite{mirzaEndtoendMultimodalStructure2026} \\
\addlinespace
Houthuijs \textit{et al.}~\cite{houthuijsSilicoInfraredSpectral2023} & computed & $>75$k spectra of $\sim$4500 metabolites & DFT, molecular ions (MS adducts) & human metabolome & IR ion spectroscopy \\
\bottomrule
\end{tabular}
\end{table}

Two structural limits of this landscape shape our work. First, every openly available DFT-quality set with more than $10^5$ molecules is either QM9-derived or \QMeS; there is no large-scale IR dataset reaching drug-like size. Second, and more important here, none of these sets contains conformational ensembles: each molecule is represented by a single minimum, so a model trained on them cannot learn to distinguish conformers from their spectra. Hessian-only datasets such as THEMol~\cite{liang2026themol} do not fill this gap either, because IR intensities also require dipole derivatives.

\subsection{Spectra from a machine-learned potential}
\label{sec:spectra_from_mlip}

Computing harmonic spectra at the DFT level for millions of structures is not affordable. Machine-learned interatomic potentials (MLIPs) that predict both forces and dipoles make it possible, because frequencies and IR intensities can then be obtained from a single model at a small fraction of the DFT cost. We therefore use one workflow for every database: geometry optimisation by BFGS line search to $\fmax = 5\times10^{-4}$~eV/\AA\ (at most 1000 steps), followed by a harmonic analysis from central finite differences of the forces (displacement 0.01~\AA), with intensities from the derivatives of the predicted dipole. Frequencies are reported unscaled throughout; no scaling factor or per-method calibration is applied.

We compared three foundation MLIPs that provide forces and dipoles, SO3LR~\cite{kabyldaMolecularSimulationsPretrained2025}, AIMNet2~\cite{anstineAIMNet2NeuralNetwork2025} and MACE4IRmol~\cite{bhatiaMACE4IRmolUncertaintyawareFoundation2026}, on two benchmarks: the frequencies of VIBFREQ1295~\cite{zapatatrujilloVIBFREQ1295NewDatabase2022}, measured against CCSD(T)(F12*) harmonic frequencies and against experimental fundamentals, and 6517 neutral closed-shell gas-phase spectra from the NIST WebBook~\cite{nistChemistryWebBook}, compared through the spectral cosine after Lorentzian broadening (15\wn FWHM, 500--4000\wn). The B3LYP/6-31G* harmonic spectra distributed by NIST serve as the DFT baseline. Details and full results are given in Section~\ref{sec:mlip_choice} (Figure~\ref{fig:mlip_benchmark}, Table~\ref{tab:mlip_benchmark}).

The three models separate along different axes. MACE4IRmol is the most accurate on frequencies, with a median absolute error against CCSD(T) of 14.3\wn against 23.2\wn for SO3LR, but on experimental spectra this advantage largely disappears: after the rigid shift that maximises the cosine, the three MLIPs lie within 0.04 cosine of each other and within 0.06 of the DFT baseline. AIMNet2 is the fastest per molecule but converged on only 2033 of the 6517 NIST molecules, which rules it out for unattended use on millions of structures. MACE4IRmol converges as reliably as SO3LR but costs about five times more serial CPU time on the benchmark, and its cost grows faster with system size (geometry optimisation scaling as $N^{2.17}$ against $N^{1.28}$ for SO3LR).

We therefore use SO3LR. It is the only model that combines complete convergence on the benchmark with a cost compatible with millions of structures, and its agreement with experiment is close to that of both the other MLIPs and the DFT baseline. Its restriction to H, C, N, O, F, P, S and Cl is the price paid, and it is a small one: those eight elements already cover almost all of our source structures.

\subsection{Starting structures}
\label{sec:starting_structures}

We generate spectra for three structure databases, each chosen for a different role. Table~\ref{tab:starting_structures} lists the role of each one and the amount of data we use from it.

\begin{table}[htb!]
\caption{The three structure databases used as a starting point, their role and the amount of data used.}
\label{tab:starting_structures}
\begin{tabular}{p{5.0cm}p{3.0cm}p{4.2cm}p{4.2cm}}
\toprule
Database & Role & Starting structures & Size used \\
\midrule
\PCQM~\cite{huOGBLSCLargeScaleChallenge2021}, the public 3D release of PubChemQC~\cite{nakataPubChemQCProjectLargeScale2017} & molecular diversity & B3LYP/6-31G* minima & 3.5~M structures, of which 3.3~M (97.5\%) contain only SO3LR elements \\
\addlinespace
\GEOM~\cite{axelrodGEOMEnergyannotatedMolecular2022} & conformational diversity & CREST/GFN2-xTB conformer ensembles of 133k QM9 molecules & 1.8~M conformers \\
\addlinespace
\QMeS~\cite{yuanQMe14SComprehensiveEfficient} & DFT fine-tuning set & B3LYP/TZVP minima & 186k molecules \\
\bottomrule
\end{tabular}
\end{table}

\PCQM supplies molecular diversity at drug-like size; it carries the original PubChemQC geometries and is the only PubChemQC release whose structures were minimised at the DFT level; later releases use PM6 geometries~\cite{nakataPubChemQCPM6Data2020}. \GEOM supplies the conformational diversity that no existing IR dataset provides, and \QMeS the DFT-level spectra for fine-tuning and for testing transfer. Every structure is re-optimised with SO3LR before the harmonic analysis, so all final geometries are SO3LR minima; for \GEOM this means several input conformers of one molecule can relax into the same minimum, which the duplicate filter below removes.

\begin{figure}[htb!]
    \centering
    \includegraphics[width=0.8\linewidth]{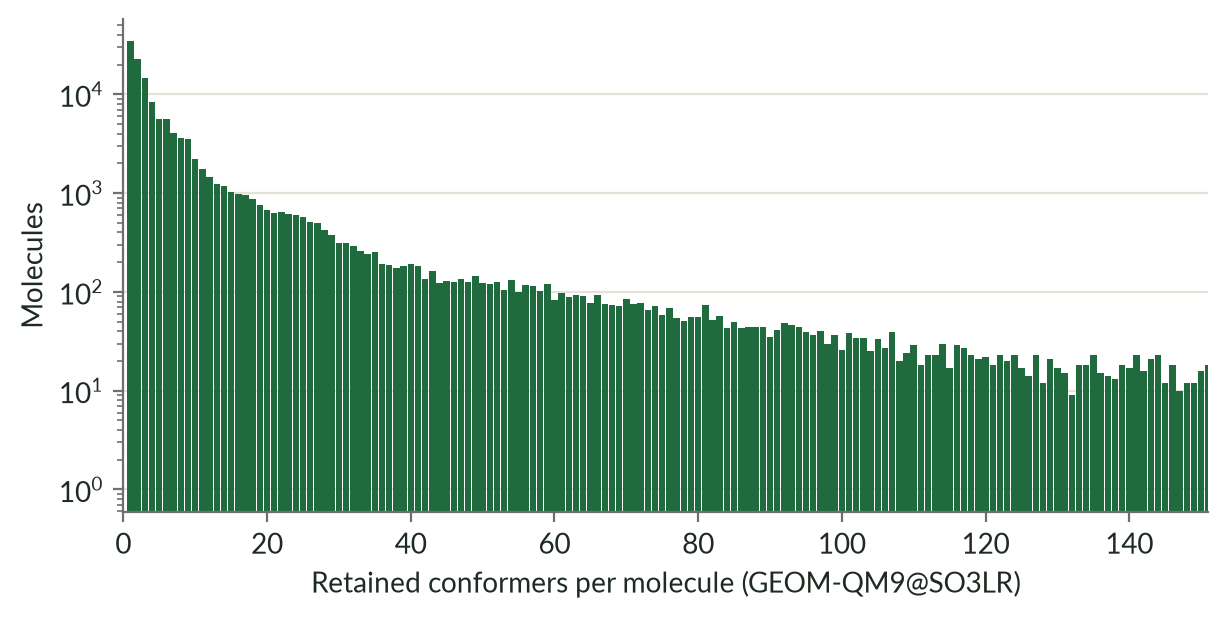}
    \caption{Retained conformers per molecule in \GEOMSO (log scale, truncated at 150).}
    \label{fig:geom_conformers}
\end{figure}

To describe the available conformational diversity, Figure~\ref{fig:geom_conformers} gives the number of retained conformers per \GEOM molecule.

\subsection{Data preparation and cleaning}
\label{sec:cleaning}

\paragraph{Elements, convergence and imaginary frequencies.} All datasets are restricted to the elements covered by SO3LR (H, C, N, O, F, P, S, Cl), including \QMeSDFT, so that every database describes the same chemistry. This keeps 92.6\% of \QMeS and 97.5\% of \PCQM; \GEOM contains only H, C, N, O and F. Every retained SO3LR structure is converged ($\fmax < 5\times10^{-4}$~eV/\AA); the few calculations that failed ($<0.2\%$ per database) are not included. We then remove every structure with at least one imaginary frequency ($\nu_\text{min} < 0$).

\paragraph{Chemical validity.} To avoid introducing chemically unreasonable structures into the training set, which could encourage the generative model to reproduce such configurations, we follow the cleaning and validation procedure proposed by Nikitin \textit{et al.}~\cite{nikitinGEOMdrugsRevisitedMore2025}. Molecules that fragment into multiple disconnected components during geometry optimisation are excluded, and chemical validity is assessed from the reconstructed molecular graph using the element-, formal-charge- and valency-based criteria defined there. Fragmentation is common in \GEOM, where 110{,}292 conformers (6.0\%) lose a small molecule, most often \ch{N2}, \ch{CO2} or \ch{NH3}, during the SO3LR relaxation, against only 708 structures in \PCQM. Most of these are already caught by the imaginary-frequency filter above; once it has run, the graph filter removes a further 19{,}782 structures from \GEOM and 13{,}348 from \PCQM.

To infer molecular connectivity from atomic coordinates, we use the extended H\"uckel theory (EHT) implementation in RDKit~\cite{kimUniversalStructureConversion2015, landrumRDKitOpenSource} rather than relying solely on distance-based bond assignment. On \QMeS, EHT reduces graph-construction failures from 9249 (connect-the-dots) and 9498 (van der Waals radii) to 2327, while preserving chemically plausible bonding patterns. Inferring molecular bonds from a set of atoms and their spatial coordinates has no universally applicable solution, in part because the concept of a chemical bond itself lacks a unique and rigorous definition~\cite{gonthierQuantificationFuzzyChemical2012}; we adopt EHT so that the validation criteria of Nikitin \textit{et al.}~\cite{nikitinGEOMdrugsRevisitedMore2025} can be applied consistently to every database.

\paragraph{\QMeSDFT spectra.} Some of the IR spectra distributed with \QMeS contain spurious zero intensities, so we recompute every spectrum from the stored Hessian and dipole derivatives. The \QMeS geometries also include some molecules %
that are already broken into disconnected fragments; the graph filter above removes them. Wu \textit{et al.} report the same kind of mismatch between geometry and SMILES for 1873 \QMeS molecules~\cite{wuLatentDiffusionbased3D2026}.

\paragraph{Duplicates.} Before splitting we remove geometric duplicates, that is, structures sharing a canonical SMILES with a retained structure and lying within a Kabsch-aligned RMSD of 0.05~\AA\ of it. This matters mostly for \GEOM, where 157{,}378 conformers (8.6\%) relax into a minimum already occupied by another conformer of the same molecule.

Together with the imaginary-frequency and graph filters applied before it, this removes 0.7\% of \QMeSDFT, 3.9\% of \PCQM and 19.2\% of \GEOM; imaginary frequencies are the largest cause in both SO3LR databases (2.8\% of \PCQM, 9.5\% of \GEOM).

\subsection{Chemical and spectral diagnostics}
\label{sec:diagnostics}

\begin{figure}[htb!]
    \centering
    \includegraphics[width=0.6\linewidth]{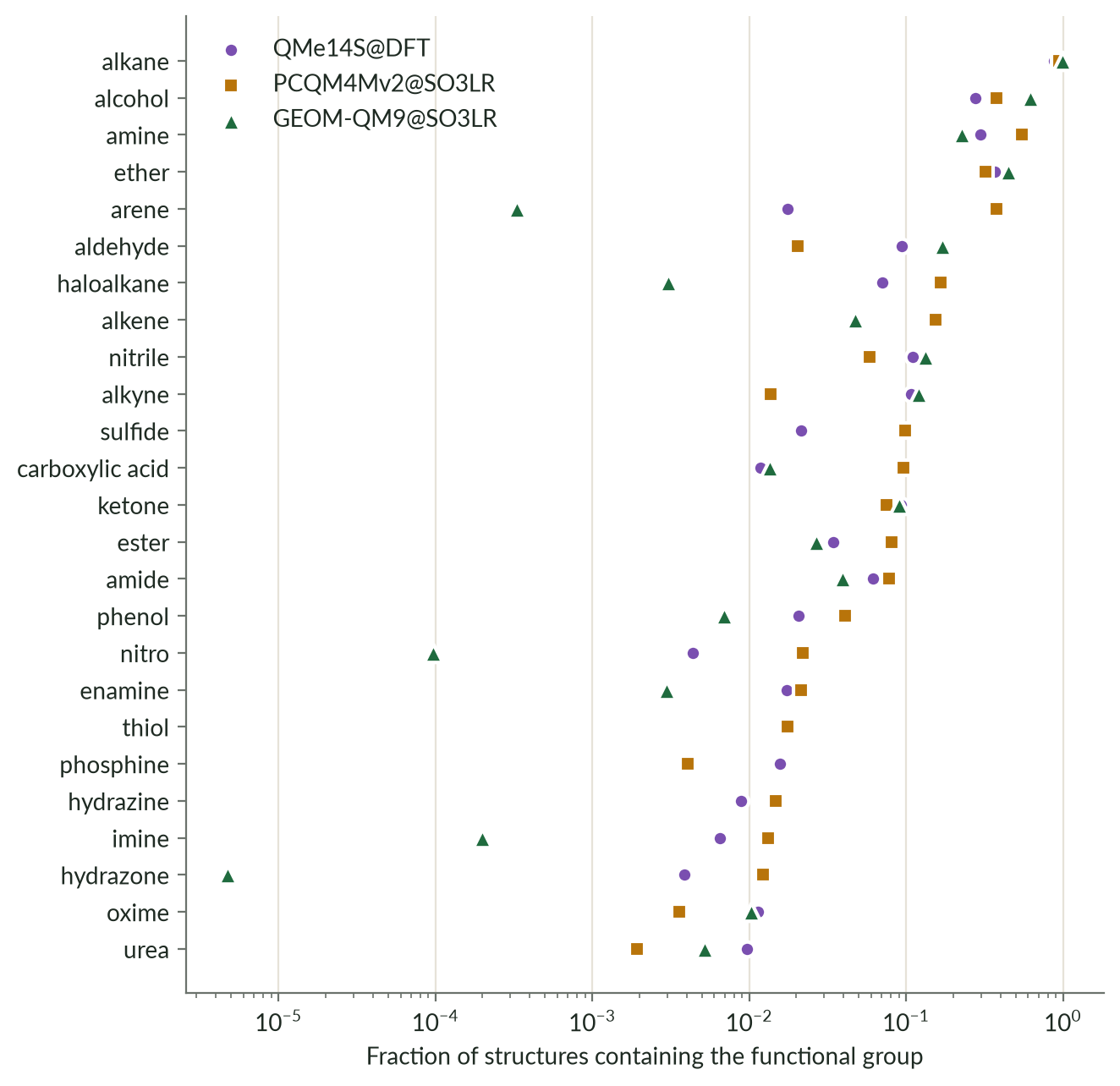}
    \caption{Fraction of structures matching each functional-group SMARTS pattern (log scale). The 25 groups shown, of 48, are those with the highest fraction in at least one of the three databases.}
    \label{fig:functional_groups}
\end{figure}

The databases differ in their chemical coverage. \QMeS and \PCQM cover all eight SO3LR elements while \GEOM contributes only H, C, N, O and F, and \PCQM supplies most of the F-, S- and Cl-containing structures in absolute numbers (304k, 441k and 253k; Figure~\ref{fig:pineirdb_qme14s_map}). To check whether the databases also differ in chemistry beyond the elements, Figure~\ref{fig:functional_groups} compares the fraction of structures carrying each functional group. \PCQM brings arenes (38\% of its structures), sulfides, carboxylic acids and nitro groups, all rare in \QMeS and rarer still in \GEOM, whereas aldehydes, nitriles and alkynes are more frequent in \QMeS and \GEOM.

\begin{figure}[htb!]
    \centering
    \includegraphics[width=0.6\linewidth]{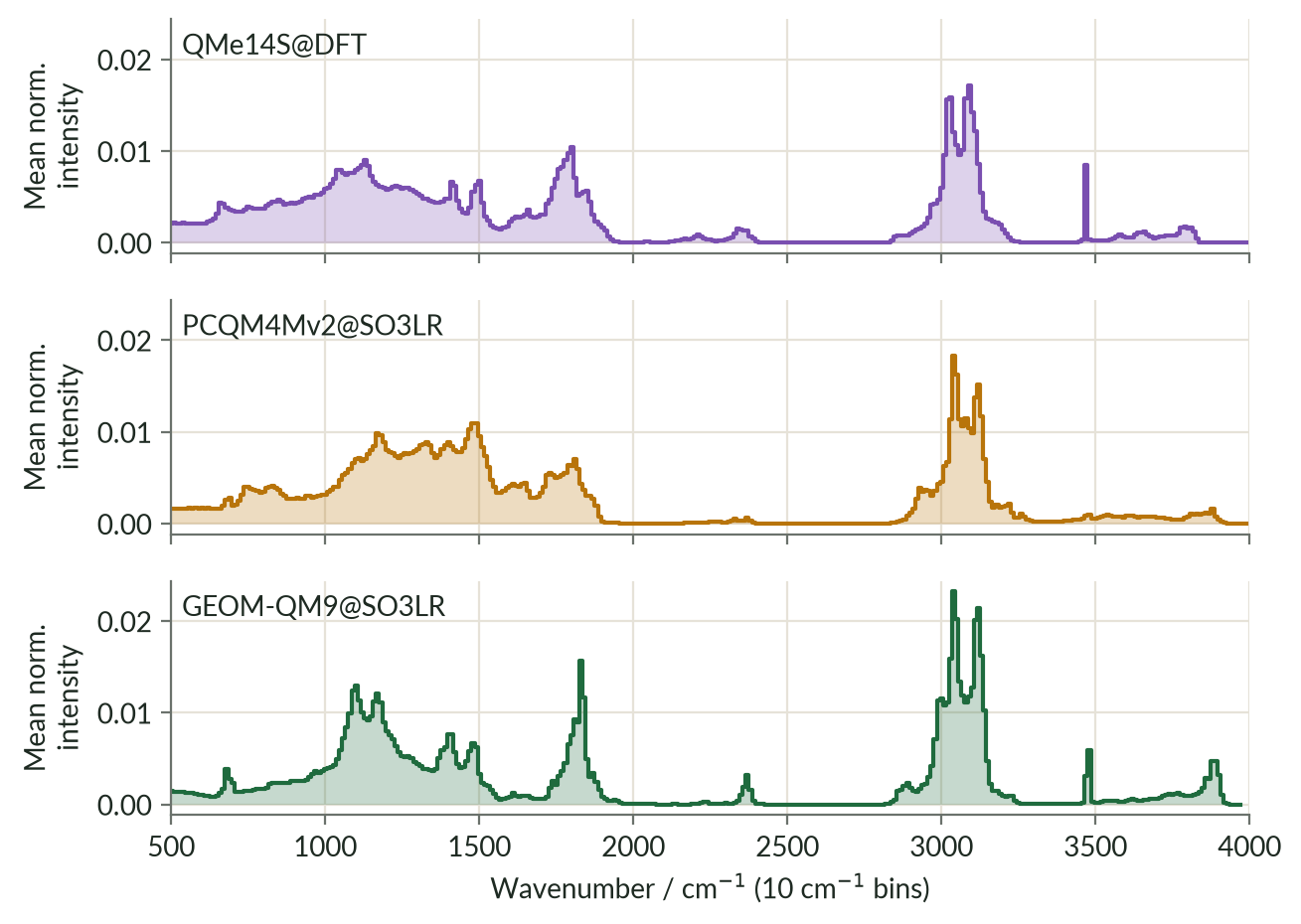}
    \caption{Mean area-normalised stick spectrum per database (10\wn bins).}
    \label{fig:mean_spectra}
\end{figure}

To see whether these differences in chemistry reach the spectra, Figure~\ref{fig:mean_spectra} compares the mean spectrum of each database. All three show the expected bands, C--H stretches around 3000\wn, the carbonyl band near 1700--1850\wn and a dense fingerprint region below 1500\wn, but they diverge where their chemistry diverges: \GEOM shows strong nitrile and alkyne bands (C$\equiv$N and C$\equiv$C near 2350\wn and $\equiv$C--H near 3480\wn, unscaled harmonic SO3LR positions), while \PCQM has a broader fingerprint region and aromatic C--H stretches above 3050\wn.

\subsection{Splitting}
\label{sec:splitting}

Each dataset is split by molecule: all structures sharing a canonical SMILES, such as the conformers of one GEOM molecule, go to the same partition. The \QMeSDFT test set holds 1000 molecules chosen to cover chemical space rather than to follow the data: they are spread evenly over molecular size and selected so that every element and most functional groups are represented. Of the remaining molecules, 5\% form the validation set and the rest the training set. To check that the test set covers chemical space as intended, Figure~\ref{fig:pineirdb_qme14s_map} (lower left) compares it with training and validation by molecular size and by element. Training and validation follow the size distribution of \QMeS, which peaks around 20 atoms and thins out beyond 30, while the test set is flat in size from 4 to 30 atoms, and F, P, S and Cl are several times more frequent in it than in training.

\QMeSSO and \PineDB reuse this partition: every molecule shared with \QMeSDFT keeps the partition it has there. This keeps the \QMeS test molecules out of training and validation in the aligned splits.

\PineDB is split into training (3.03~M molecules) and validation (57k molecules). After removing the \QMeS test molecules, 1000 further \PineDB molecules are drawn at random (seed 0) as \PineDB-test: 901 from \PCQM and 99 from \GEOM, with 2360 structures in all. The generative benchmark scores one structure per molecule. \PineDB reaches larger molecules than \QMeS, peaking near 30 atoms with a tail to 53 atoms (Figure~\ref{fig:pineirdb_qme14s_map}).

Models are tested on two versions of the same test split: \QMeSDFT-test (1000 molecules) at the DFT level, and \QMeSSO-test at the SO3LR level, where 919 of the molecules pass SO3LR re-optimisation and the cleaning filters. The \PineDB structures of the 510 \QMeS test molecules that \PineDB contains form the \QMeS conformer set. Its \GEOM part, 356 molecules with 9051 conformers (a median of 3 per molecule), lets predictions be tested across the conformers of one molecule. Figure~\ref{fig:cleaning_workflow} summarises the cleaning and splitting workflow.

\subsection{Joint view of spectral coverage and composition}
\label{sec:coverage}

\begin{figure}[htb!]
    \centering
    \includegraphics[width=0.85\linewidth]{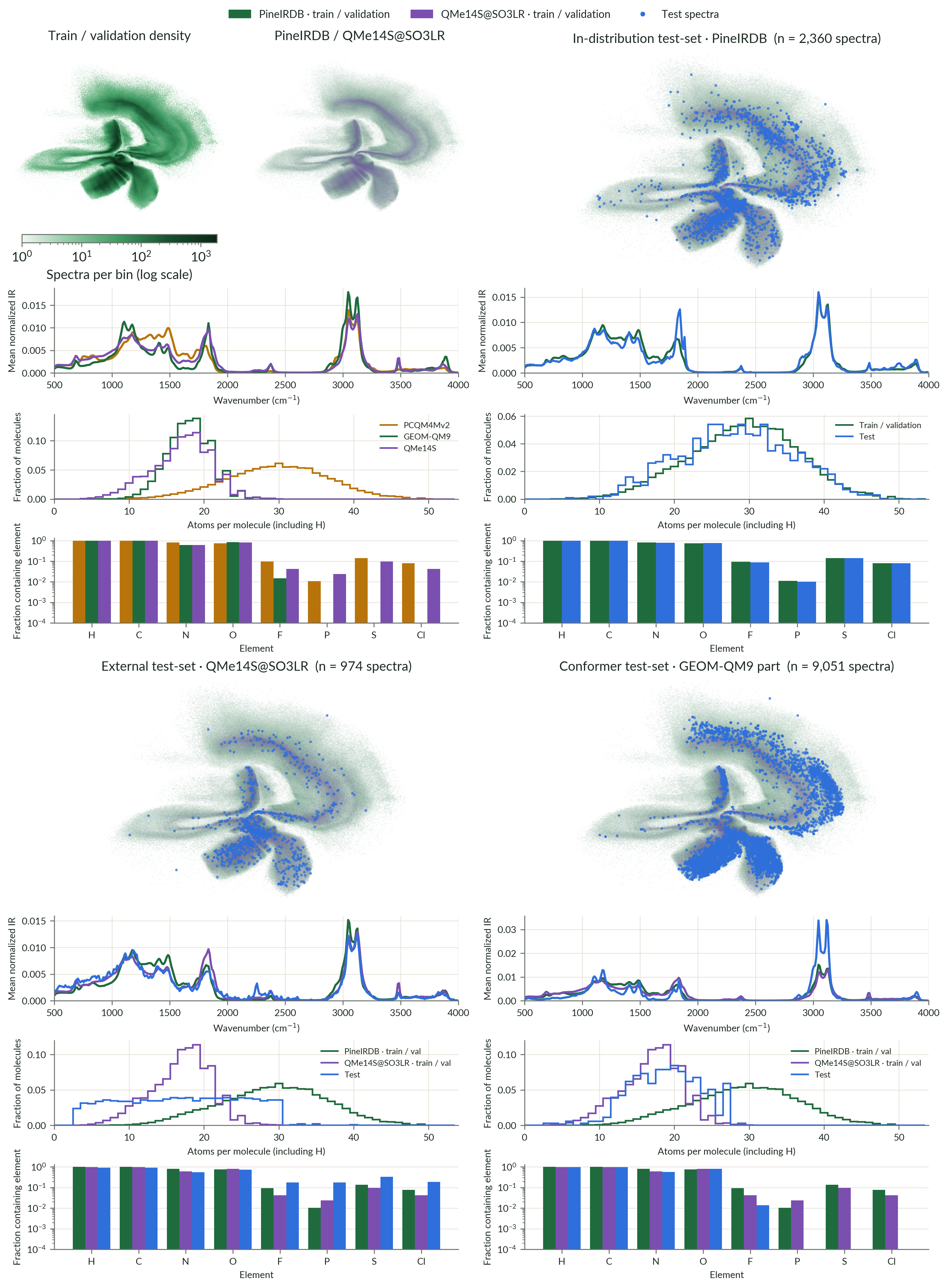}
    \caption[Spectral and chemical coverage of \PineDB and \QMeSSO.]{Spectral and chemical coverage of \PineDB and \QMeSSO. Upper left: combined training/validation spectral density on a logarithmic scale and the dataset overlay (\PineDB green, \QMeSSO purple). The other map panels mark test spectra in blue: \PineDB-test (upper right, 2360 spectra belonging to the 1000 test molecules), \QMeSSO (lower left, 974 spectra), and the \GEOM part of the \QMeS conformer set (lower right, 9051 spectra). The maps use the same two-dimensional spectral representation as Figure~{3} of the main text. Each group of panels also compares mean normalised spectra, atom counts including hydrogen, and the fraction of molecules containing each element. In the source-dataset comparisons, \PCQM is amber, \GEOM green and \QMeS purple; the test comparisons use blue for the test subset. Mean spectra weight structures equally, including individual conformers, and use 10\wn bins followed by 15\wn Lorentzian broadening; size and element summaries count distinct molecules. The plotted \GEOM population is not identical to the scored multi-conformer subset in Table~{IV} of the main text (9085 spectra). All spectra in this figure are computed at the SO3LR level.}
    \label{fig:pineirdb_qme14s_map}
\end{figure}

To compare the evaluation sets with the spectral and chemical coverage of the training data, Figure~\ref{fig:pineirdb_qme14s_map} places them on the same two-dimensional spectrum map and compares their mean spectra, molecular sizes and elemental composition. \PineDB-test broadly follows the training distribution, whereas \QMeS spans small molecules and rare elements more evenly; the \GEOM part of the \QMeS conformer set concentrates on smaller molecules and a narrower elemental range. Overlap on this compressed map describes coarse spectral coverage, not molecular identity or evidence that a test molecule appeared in training.

\clearpage
\section{Evaluation sets, protocols and full results}
\label{sec:evaluation}

This section collects the definitions and the complete tables behind the results of the main text. The main text reports the rates of correct SMILES, while the tables below also give the matched RMSD and the fraction of chemically stable candidates.

\subsection{Evaluation sets}
\label{sec:evalsets}

We use four evaluation sets, listed in Table~\ref{tab:evalsets}. They differ in chemical distribution, in level of theory and in what they measure.

\begin{table}[htb!]
\caption{The four evaluation sets. The name in the first column is the one used in the main text.}
\label{tab:evalsets}
\footnotesize
\renewcommand{\arraystretch}{1.25}
\begin{tabularx}{\textwidth}{@{}llllX@{}}
\toprule
Name & Database & Spectra & Size & What it measures \\
\midrule
External test set & \QMeSSO & SO3LR & 974 spectra, 919 molecules & correct SMILES for held-out molecules of a separately curated database, at the training level of theory \\
In-distribution test set & \PineDB-test & SO3LR & 1000 molecules & correct SMILES within the chemical distribution of the training database \\
Conformer test set & \QMeS conformer set & SO3LR & 9085 spectra, 291 molecules & whether the generated geometry reproduces the conformer of the input spectrum \\
DFT test set & \QMeSDFT test split & B3LYP/TZVP & 1000 molecules & correct SMILES for the molecules of the external test set at a more accurate level of theory, after fine-tuning \\
\bottomrule
\end{tabularx}
\end{table}

\paragraph{External test set.} We split \QMeS by molecule, with a test partition of 1000 molecules chosen to cover size, elements and functional groups (Section~\ref{sec:splitting}). Every \QMeS test molecule is held out of \PineDB training by canonical SMILES. After the SO3LR re-optimisation 919 of these molecules remain, with 974 of the 1062 test structures. The other molecules fail the imaginary-frequency or the graph checks. The generative benchmark scores all 974 structures, and the forward-model benchmark the 969 structures that are IR-active, so the two means are not taken over identical sets.

\paragraph{In-distribution test set.} After removing the \QMeS test molecules, we drew 1000 \PineDB molecules at random and held them out of training, 901 from \PCQM and 99 from \GEOM. Since they are drawn at random and not chosen for coverage, they follow the training distribution.

\paragraph{Conformer test set.} The \QMeS conformer set contains the \PineDB structures of the 510 \QMeS test molecules that are also in \PineDB (9304 spectra). These molecules are held out of training and validation like the rest of the \QMeS test set, but their spectra are the SO3LR spectra of the \PineDB conformers, and not of the single \QMeS geometry. The conformer analysis uses the 291 molecules with at least two retained conformers (9085 spectra), and the diastereomer analysis uses all 510 molecules. This set is distinct from the in-distribution test set.

\paragraph{DFT test set.} This is the same \QMeS test partition with its original B3LYP/TZVP spectra. It contains all 1000 molecules, including the 81 molecules removed by the SO3LR re-optimisation. It tests the molecules of the external test set at a different and more accurate level of theory. Note that it is not external for the models fine-tuned on \QMeSDFT, since these models are trained on the other molecules of the same database.

\subsection{Metrics}
\label{sec:metrics}

\paragraph{Chemical stability.} A generated structure is chemically stable when the bond graph inferred from its coordinates is connected and satisfies the valence and formal-charge criteria of Section~\ref{sec:cleaning}, which are also applied to the training data. This criterion corresponds to the molecular stability of EDM~\cite{pmlr-v162-hoogeboom22a} and KRONOS~\cite{ottomano2026autoregressive}, with connectivity added, and is stricter than their 3D validity. Candidates that are not chemically stable count as failures.

\paragraph{Correct SMILES.} A candidate is correct when it is chemically stable and its canonical SMILES, without stereochemistry, equals that of the reference. We ignore stereochemistry because enantiomers have the same IR spectrum. Diastereomers are tested separately in the main text. ``Correct at the first attempt'' and ``correct within 50 attempts'' are the fractions of test spectra with a correct candidate among the first one and the first 50 candidates, in generation order, with chemically stable candidates placed first. ``Correct best-ranked SMILES'' is the fraction of spectra whose candidate with the highest spectral cosine is correct, and is used only in the ranking analysis.

\paragraph{Geometry.} The geometry of a correct candidate is compared with the reference through the root-mean-square deviation
\begin{equation}
\mathrm{RMSD} = \min_{\pi,\,O} \sqrt{\frac{1}{N} \sum_{i=1}^{N} \bigl\lVert O\, \hat{\mathbf{x}}_{\pi(i)} - \mathbf{x}_i \bigr\rVert^2} ,
\end{equation}
where $\mathbf{x}_i$ and $\hat{\mathbf{x}}_i$ are the centred positions of atom $i$ in the reference and in the candidate, $N$ is the number of atoms, $\pi$ is a mapping between the two molecular graphs, and $O$ is the orthogonal matrix of the Kabsch alignment~\cite{kabschSolutionBestRotation1976}, with reflections allowed. The RMSD is computed only for correct candidates, so for each model it is averaged over a different subset of molecules.

\subsection{Protocols}
\label{sec:protocols}

\paragraph{Fixed number of attempts.} In Tables~\ref{tab:full_external}, \ref{tab:full_indist} and~\ref{tab:full_dft}, every model generates 50 attempts per spectrum. Chemically unstable structures remain failures and are not replaced. Chemically stable candidates are placed first, and the generation order is kept within the stable and the unstable groups. The spectral cosine of \PineIR is not used in these comparisons.

\paragraph{Extra input imposed on \PineIR.} The \PineIR rows with extra input use the same checkpoint as the row without it, and no retraining is involved. With the atom count, the stop token is forbidden until the given number of atoms is reached. With the formula, only elements of the formula that are not yet used up can be generated (Section~\ref{sec:ar}). In both cases generation is forced to continue after the point where the model would stop, which lowers the fraction of chemically stable candidates.

\paragraph{Ranking.} The ranking analysis of the main text uses a separate protocol. For each spectrum we collect the first 50 chemically stable candidates, with at most 500 attempts. Subsets of these candidates show how the rate of correct best-ranked SMILES changes with the size of the candidate pool. Since the number of attempts differs, these values cannot be compared directly with the fixed 50-attempt benchmark. The best candidate of a list, or the candidate with the highest structural similarity to the reference, can only be identified with the reference structure. These quantities measure the quality of the list, and not a selection that the model can make on its own.

\paragraph{Conformers.} For the molecules of the conformer test set, each spectrum is paired with up to 10 other conformers of the same molecule that differ from it by more than 0.1~\AA\ in RMSD. Every correct candidate is compared with the conformer that produced the spectrum and with these alternatives. The analysis is conditional on a correct SMILES and is averaged over molecules. Candidates with a wrong SMILES are not counted as successful conformer predictions.

\paragraph{Diastereomers.} We determine the stereochemistry of each correct candidate and of the reference from their 3D coordinates, and count mirror images as one. We report the fraction of correct candidates with the diastereomer of the reference, on the 136 molecules of the conformer test set with at least two diastereomers for which all three models generate the correct SMILES, and on the 91 such molecules of the external test set.

\paragraph{Baselines.} DiffSpectra~\cite{wangDiffSpectraMolecularStructure2025v2} receives the IR spectrum and the atom count, and IR-GeoDiff~\cite{wuLatentDiffusionbased3D2026} the IR spectrum and the molecular formula. In the main comparisons both are trained on the same \PineDB split at the SO3LR level as \PineIR. The baselines therefore receive more information than \PineIR, and the comparison tests whether this information is needed, rather than isolating the effect of the architecture. The published checkpoints, trained on QM9S, are discussed in Section~\ref{sec:published_checkpoints}.

\subsection{Full results}
\label{sec:full_results}

Tables~\ref{tab:full_external} and~\ref{tab:full_indist} extend Table~{III} of the main text with the matched RMSD and the fraction of chemically stable candidates. Table~\ref{tab:full_conformers} extends Table~{IV} with the fraction of molecules for which a model is closer to the input conformer in more than half of the comparisons, and Table~\ref{tab:full_dft} extends Table~{V}. DiffSpectra generates a larger fraction of chemically stable structures than \PineIR throughout. This fraction is the denominator of the rates of correct SMILES, and we report it as a diagnostic and not as a measure of the quality of the elucidation.

\begin{table}[htb!]
\caption{Full results on the external test set, with 50 attempts per spectrum, grouped by the information that each model receives in addition to the spectrum. All models are trained on the same \PineDB SO3LR split. The matched RMSD is averaged over the correct candidates of each model, so the underlying molecules differ between rows. No spectral ranking is used. The best value of each group is in bold.}
\label{tab:full_external}
\begin{tabular}{lrrrr}
\toprule
Model & Correct at first & Correct within & Matched & Chemically \\
      & attempt (\%) $\uparrow$ & 50 attempts (\%) $\uparrow$ & RMSD (\AA) $\downarrow$ & stable (\%) \\
\midrule
\multicolumn{5}{@{}l}{\textit{Spectrum only}} \\
\PineIR     & 31 & 57 & 0.59 & 72 \\
\addlinespace
\multicolumn{5}{@{}l}{\textit{Spectrum and atom count}} \\
\PineIR     & \textbf{37} & 58 & 0.59 & 50 \\
DiffSpectra & 22 & \textbf{63} & 0.59 & 85 \\
\addlinespace
\multicolumn{5}{@{}l}{\textit{Spectrum and molecular formula}} \\
\PineIR     & \textbf{45} & 72 & 0.58 & 57 \\
IR-GeoDiff  & 25 & \textbf{73} & 0.72 & 92 \\
\bottomrule
\end{tabular}
\end{table}

\begin{table}[htb!]
\caption{Full results on the in-distribution test set, grouped and scored as in Table~\ref{tab:full_external}, with the same \PineIR checkpoint and the same 50 attempts per spectrum.}
\label{tab:full_indist}
\begin{tabular}{lrrrr}
\toprule
Model & Correct at first & Correct within & Matched & Chemically \\
      & attempt (\%) $\uparrow$ & 50 attempts (\%) $\uparrow$ & RMSD (\AA) $\downarrow$ & stable (\%) \\
\midrule
\multicolumn{5}{@{}l}{\textit{Spectrum only}} \\
\PineIR     & 41 & 79 & 0.77 & 81 \\
\addlinespace
\multicolumn{5}{@{}l}{\textit{Spectrum and atom count}} \\
\PineIR     & \textbf{51} & \textbf{79} & 0.77 & 55 \\
DiffSpectra & 22 & 75 & 0.96 & 95 \\
\addlinespace
\multicolumn{5}{@{}l}{\textit{Spectrum and molecular formula}} \\
\PineIR     & \textbf{50} & \textbf{80} & 0.77 & 59 \\
IR-GeoDiff  & 11 & 61 & 1.09 & 91 \\
\bottomrule
\end{tabular}
\end{table}

\begin{table}[htb!]
\caption{Full conformer sensitivity on the conformer test set. The protocol is described in Section~\ref{sec:protocols}. ``Closer'' is the percentage of comparisons in which a correct candidate is closer to the conformer that produced the spectrum, and a model that ignores the 3D geometry gives 50\%. ``Molecules above chance'' is the percentage of molecules for which this value exceeds 50\%. Each model is evaluated on the molecules for which it generates the correct SMILES (267, 265 and 257 of 291), so the underlying molecules differ between rows. Means are over molecules, with 95\% bootstrap confidence intervals in brackets.}
\label{tab:full_conformers}
\begin{tabular}{lcccc}
\toprule
Model & RMSD to the input & RMSD to another & Closer & Molecules above \\
 & conformer (\AA) $\downarrow$ & conformer (\AA) & (\%) $\uparrow$ & chance (\%) $\uparrow$ \\
\midrule
DiffSpectra (IR and atom count) & 0.73 [0.69, 0.78] & 0.98 [0.93, 1.01] & 68 [66, 70] & 93 \\
IR-GeoDiff (IR and formula)     & 0.96 [0.92, 0.99] & 1.04 [1.00, 1.08] & 58 [56, 59] & 80 \\
\PineIR (IR)                    & 0.58 [0.54, 0.61] & 0.99 [0.95, 1.03] & 78 [76, 81] & 95 \\
\bottomrule
\end{tabular}
\end{table}

\begin{table}[htb!]
\caption{Full results on the DFT test set, with 50 attempts per spectrum, scored as in Table~\ref{tab:full_external}. Models marked \PineDB $\rightarrow$ \QMeSDFT are pretrained on the \PineDB SO3LR spectra and fine-tuned on the \QMeSDFT training split. The model marked ``\QMeSDFT only'' sees no SO3LR spectra at any stage, and its autoencoder shares only the structure-only \PCQM pretraining of all \PineIR models. All models are trained and tested on unscaled frequencies. The published QM9S checkpoints of DiffSpectra and IR-GeoDiff are reported in Table~\ref{tab:published_checkpoints}.}
\label{tab:full_dft}
\footnotesize
\begin{tabular}{llrrrr}
\toprule
Model (inputs) & Training data & Correct at first & Correct within & Matched & Chemically \\
 & & attempt (\%) $\uparrow$ & 50 attempts (\%) $\uparrow$ & RMSD (\AA) $\downarrow$ & stable (\%) \\
\midrule
\multicolumn{6}{@{}l}{\textit{All test molecules (1000)}} \\
DiffSpectra (IR and atom count) & \PineDB $\rightarrow$ \QMeSDFT & \textbf{34} & \textbf{77} & 0.51 & 86 \\
\PineIR (IR) & \QMeSDFT only                  & 23 & 34 & 0.42 & 72 \\
\PineIR (IR) & \PineDB $\rightarrow$ \QMeSDFT & 32 & 46 & 0.49 & 75 \\
\addlinespace
\multicolumn{6}{@{}l}{\textit{HCNOF test molecules (527)}} \\
DiffSpectra (IR and atom count) & \PineDB $\rightarrow$ \QMeSDFT & 43 & \textbf{90} & 0.43 & 94 \\
\PineIR (IR) & \QMeSDFT only                  & 36 & 51 & 0.37 & 80 \\
\PineIR (IR) & \PineDB $\rightarrow$ \QMeSDFT & \textbf{50} & 66 & 0.36 & 86 \\
\bottomrule
\end{tabular}
\end{table}

\clearpage
\section{Choosing an MLIP for IR spectra}
\label{sec:mlip_choice}

\paragraph{Candidates.} To select an MLIP for large-scale spectral calculations, Table~\ref{tab:mlip_candidates} compares three models that provide both forces and dipoles.

\begin{table}[htb!]
\caption{The three candidate machine-learned interatomic potentials.}
\label{tab:mlip_candidates}
\begin{tabular}{lrll}
\toprule
MLIP & Parameters & Training data & Elements \\
\midrule
SO3LR~\cite{kabyldaMolecularSimulationsPretrained2025}        & 0.53~M  & 3.8~M PBE0+MBD configurations & H, C, N, O, F, P, S, Cl \\
AIMNet2~\cite{anstineAIMNet2NeuralNetwork2025}                & 2.21~M  & 20~M $\omega$B97M-D3/def2-TZVPP configurations & 14 main-group elements \\
MACE4IRmol~\cite{bhatiaMACE4IRmolUncertaintyawareFoundation2026} & 32.56~M & 10~M PBE0 configurations (QCML) & broad; covers every element of \\
 & & & the NIST benchmark \\
\bottomrule
\end{tabular}
\end{table}

\paragraph{Protocol.} All three run through the same optimisation and harmonic-analysis workflow detailed in Section~\ref{sec:spectra_from_mlip}: BFGS geometry optimisation to $\fmax = 5\times10^{-4}$~eV/\AA\ (at most 1000 steps), then a harmonic analysis by central finite differences of the forces (displacement 0.01~\AA), with intensities from the derivatives of the predicted dipole. Frequencies are unscaled. The models are scored on two benchmarks:

\begin{itemize}
\item \textbf{VIBFREQ1295}~\cite{zapatatrujilloVIBFREQ1295NewDatabase2022}: frequencies of 141 small molecules against CCSD(T)(F12*) harmonic frequencies and against experimental fundamentals. Modes are paired in sorted order after removing the rigid-body modes.
\item \textbf{NIST gas-phase spectra}~\cite{nistChemistryWebBook}: 6517 neutral closed-shell molecules. Sticks are broadened with a 15\wn FWHM Lorentzian on a 500--4000\wn grid, and the spectral cosine to the experimental spectrum is computed both directly and after the rigid shift ($\pm200$\wn) that maximises it. The B3LYP/6-31G* harmonic spectra distributed by NIST are the DFT baseline.
\end{itemize}

\begin{figure}[htb!]
    \centering
    \includegraphics[width=1.0\linewidth]{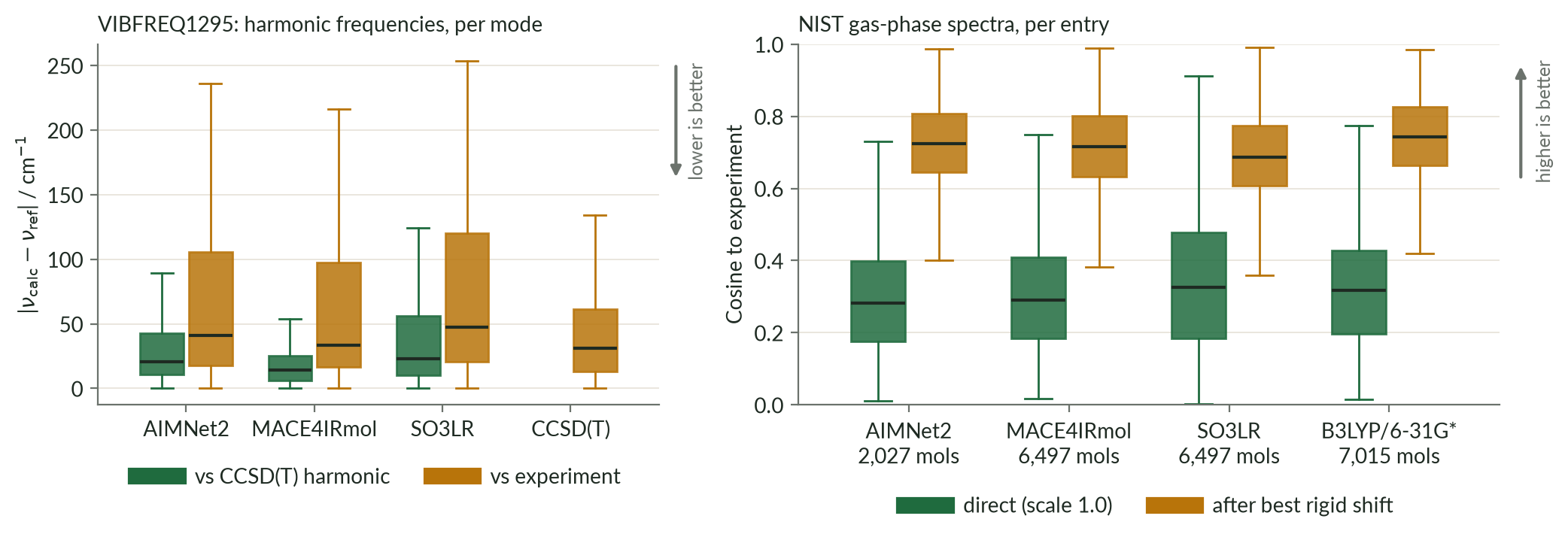}
    \caption{Left: absolute frequency error per mode on VIBFREQ1295, against CCSD(T) harmonic frequencies and against experiment (CCSD(T) itself shown for reference). Right: cosine between computed and experimental NIST spectra, before and after the best rigid shift. Boxes: interquartile range and median; whiskers: 1.5 times the interquartile range.}
    \label{fig:mlip_benchmark}
\end{figure}

To compare frequency errors with agreement to measured spectra, Figure~\ref{fig:mlip_benchmark} summarises both benchmarks; Table~\ref{tab:mlip_benchmark} reports the numerical results.

\begin{table}[htb!]
\caption{Benchmark results for the three candidate MLIPs and the DFT baseline. $^{1}$~NIST distributes the B3LYP/6-31G* baseline for a larger set of molecules than our neutral closed-shell selection. $^{2}$~Runtime model fitted on single-worker CPU timings, applied to the 8449 benchmark molecules each model supports (7708 for SO3LR).}
\label{tab:mlip_benchmark}
\begin{tabular}{lrrrr}
\toprule
 & SO3LR & AIMNet2 & MACE4IRmol & B3LYP/6-31G* \\
\midrule
VIBFREQ1295 molecules computed (of 141)                    & 116        & 112          & 116        & --- \\
median $|\Delta\nu|$ vs CCSD(T) /\wn                       & 23.2       & 20.9         & \textbf{14.3} & --- \\
mean absolute error vs CCSD(T) /\wn                                        & 53.2       & 33.1         & \textbf{21.9} & --- \\
median $|\Delta\nu|$ vs experiment /\wn (CCSD(T): 31.2)    & 47.6       & 41.0         & \textbf{33.4} & --- \\
NIST molecules computed (of 6517)                       & \textbf{6517} & 2033 & \textbf{6517} & 7015$^{1}$ \\
NIST median cosine, direct                                 & \textbf{0.327} & 0.283    & 0.290      & 0.318 \\
NIST median cosine, best shift                             & 0.686      & 0.723        & 0.716      & \textbf{0.743} \\
NIST median best shift /\wn                                & $-51$      & $-64$        & $-68$      & $-61$ \\
NIST molecules with imaginary modes                        & 511 (7.8\%) & 93 (4.6\%)  & 339 (5.2\%) & --- \\
estimated serial time, NIST benchmark$^{2}$                & 4~d 20~h   & \textbf{1~d 18~h} & 25~d 6~h & --- \\
\bottomrule
\end{tabular}
\end{table}

\paragraph{Accuracy.} MACE4IRmol is the most accurate model for frequencies, with a median error against CCSD(T) about 40\% below that of SO3LR (14.3 against 23.2\wn). On experimental spectra the differences shrink: after the best rigid shift the three MLIPs lie within 0.04 cosine of each other and within 0.06 of B3LYP/6-31G*. SO3LR has the highest direct cosine, because its harmonic frequencies are less blue-shifted (median best shift $-51$\wn against $-61$ to $-68$\wn), but the lowest shifted cosine, and it has the broadest error tail: 118 NIST spectra reach the $\pm200$\wn search limit and 7.8\% of its minima carry at least one imaginary mode.

\paragraph{Convergence.} SO3LR and MACE4IRmol converge on all 6517 NIST molecules. AIMNet2 produced results for only 2033, and already failed to converge on 16 VIBFREQ1295 molecules, so it cannot be used unattended on millions of structures.

\paragraph{Throughput.} On the NIST benchmark MACE4IRmol needs about five times the serial CPU time of SO3LR, and its cost grows faster with system size: geometry optimisation scales as $N^{2.17}$ against $N^{1.28}$ for SO3LR. AIMNet2 is the fastest per molecule, but its convergence failures cancel the advantage.

\clearpage
\section{Structure-elucidation examples}
\label{sec:examples}

\begin{figure}[htb!]
    \centering
    \includegraphics[width=0.6\linewidth]{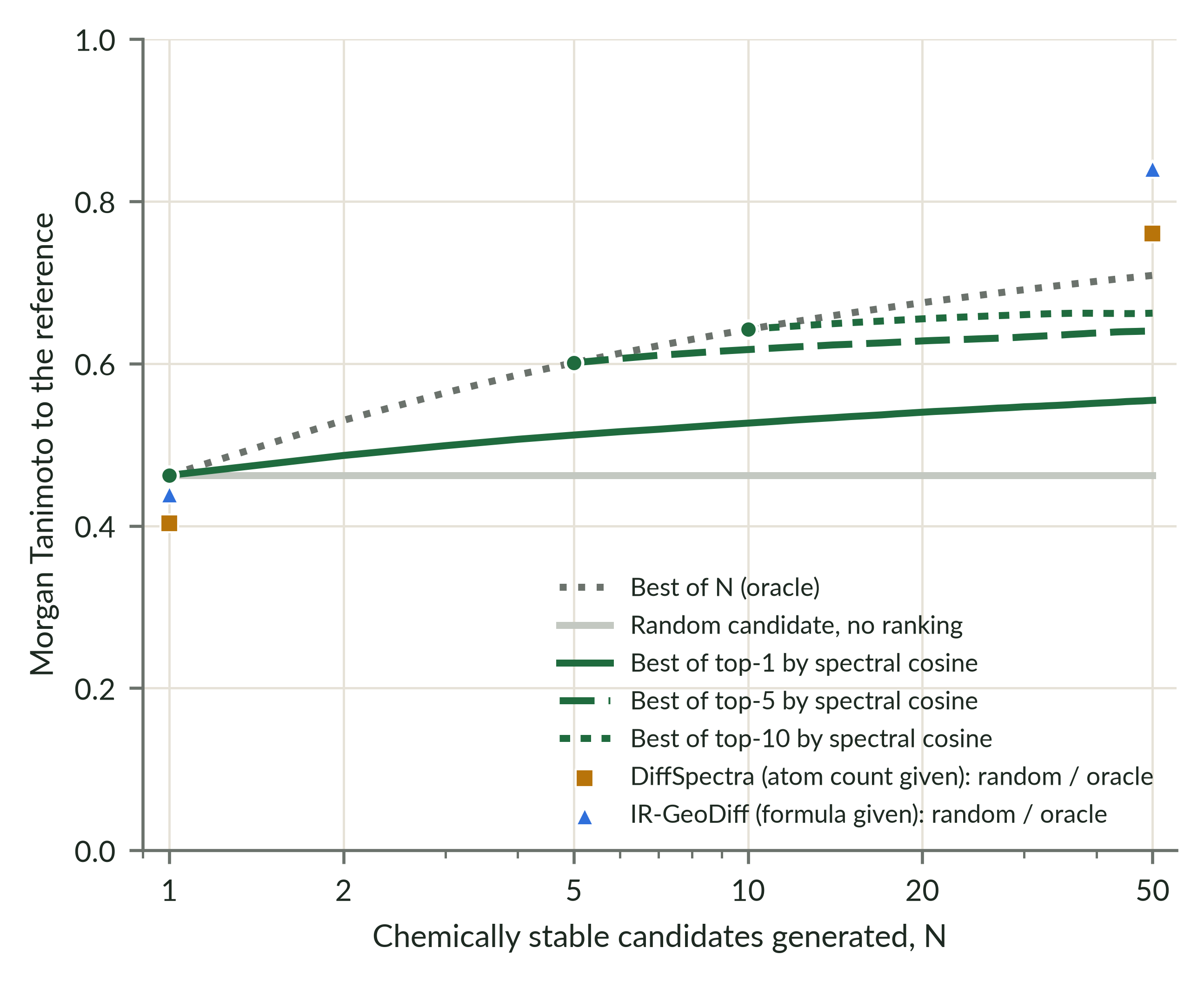}
    \caption[Tanimoto similarity of the best among the first $k$ ranked candidates.]{Morgan Tanimoto similarity to the reference of the best among the first 1, 5 and 10 candidates by spectral cosine (each curve starts at $N = k$), against the number $N$ of chemically stable candidates generated per spectrum, for \PineIR on the external test set; a random candidate (no ranking) and the best of $N$ for reference, with the random and best-of-50 values of DiffSpectra, which receives the atom count, and IR-GeoDiff, which receives the molecular formula, as markers. Curves are expected values over random subsets of 50 chemically stable candidates per spectrum.}
    \label{fig:ranking_tanimoto}
\end{figure}

To show how close the shortlist comes to the reference beyond the top-ranked candidate, Figure~\ref{fig:ranking_tanimoto} follows the Morgan Tanimoto similarity of the best among the first 1, 5 and 10 candidates by spectral cosine. This best entry can only be picked with the reference structure, so the curves measure the quality of the list, not a selection the model makes.

\begin{figure}[htb!]
    \centering
    \includegraphics[width=0.9\linewidth]{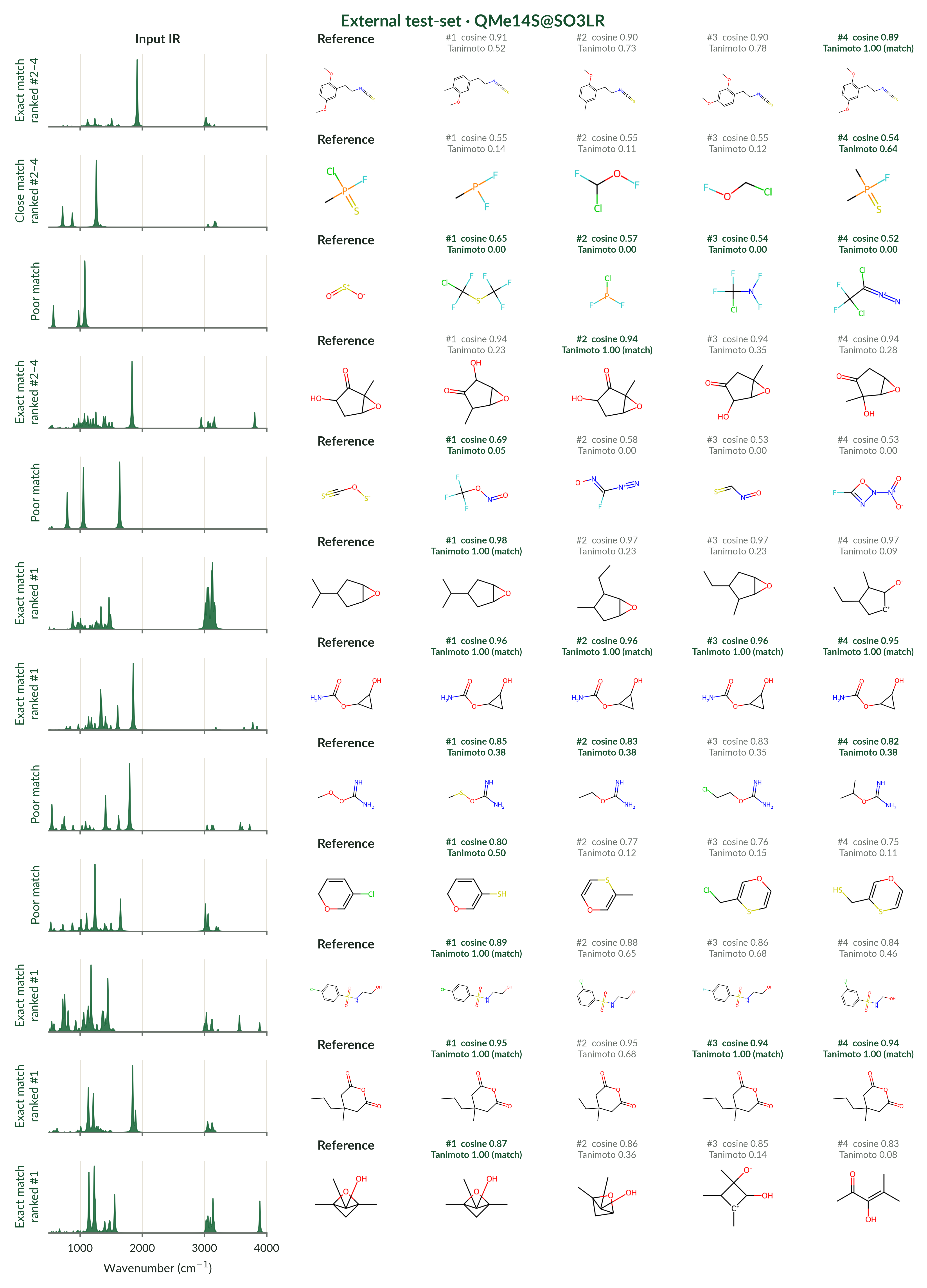}
    \caption[\PineIR on twelve random molecules of the external test set.]{\PineIR on twelve molecules of the external test set drawn at random, with no filter on size, charge or chemistry; 50 chemically stable candidates per spectrum. Each row shows the input IR spectrum (SO3LR harmonic, Lorentzian broadening of 15\wn FWHM, normalised to the highest peak), the reference structure, and \PineIR's first four distinct candidates ranked by its own IR cosine; a row is filled with repeats when the candidates hold fewer than four distinct structures. Above each candidate are its IR cosine and its Morgan Tanimoto similarity to the reference; the best candidate of the row is in bold. The row label takes the best of the four: exact match (same SMILES as the reference) ranked first or second to fourth, close match (Tanimoto $\ge 0.6$) ranked first or second to fourth, or poor match. Structures share one scale within a row, not across rows.}
    \label{fig:examples_qme14s}
\end{figure}

To show predictions without selecting favourable molecules, Figure~\ref{fig:examples_qme14s} gives twelve random examples from the external test set.

\begin{figure}[htb!]
    \centering
    \includegraphics[width=0.9\linewidth]{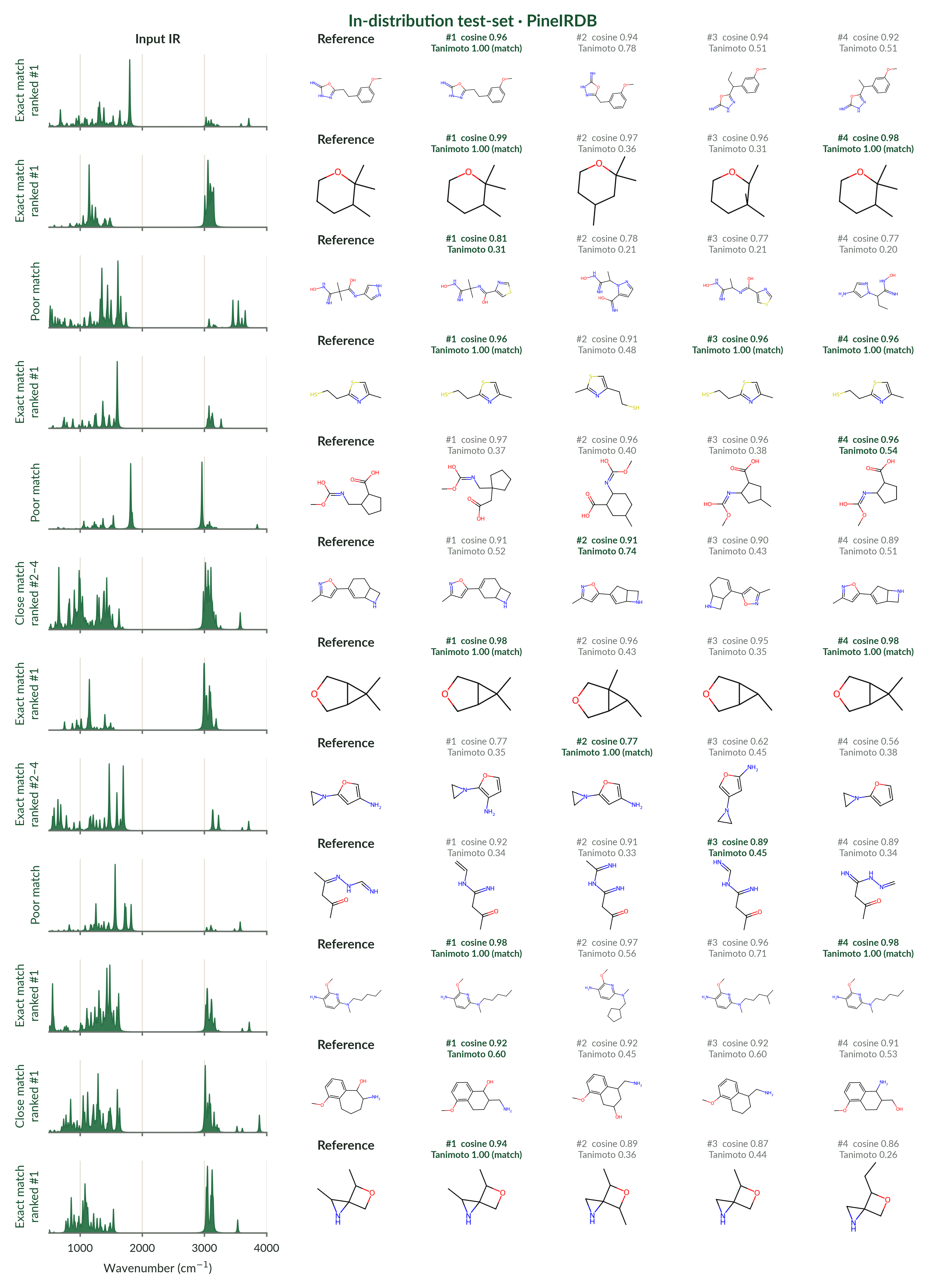}
    \caption[\PineIR on twelve random \PineDB-test molecules.]{As Figure~\ref{fig:examples_qme14s}, for twelve molecules drawn at random from \PineDB-test, 1000 \PineDB molecules held out of training (in distribution), under the fixed 50-attempt protocol rather than the 50 chemically stable candidates of Figure~\ref{fig:examples_qme14s}: \PineIR generates 50 candidates per spectrum and ranks only the chemically stable ones.}
    \label{fig:examples_pineirdb}
\end{figure}

To show the same task within the training database's distribution, Figure~\ref{fig:examples_pineirdb} gives twelve random examples from the in-distribution test set (\PineDB-test).

\clearpage
\section{Published checkpoints on the DFT test set}
\label{sec:published_checkpoints}

The published DiffSpectra and IR-GeoDiff checkpoints are trained on QM9S and support only H, C, N, O and F, so they can be run on the 527 HCNOF molecules of the \QMeSDFT test split. Their scores are reported here rather than in the main text because the comparison confounds two things at once. About 331 of those 527 molecules are QM9 molecules the published checkpoints were likely trained on, so their test is not molecule-disjoint, and IR-GeoDiff additionally receives the molecular formula while \PineIR receives none. The published checkpoints were trained on QM9S frequencies scaled by 0.965, so their input spectra are scaled by the same factor; the models we train ourselves use unscaled frequencies. Our own fine-tuned \PineIR is repeated from Table~{V} as a reference row.

\begin{table}[htb!]
\caption{Published checkpoints on the 527 HCNOF molecules of the \QMeSDFT test split, 50 attempts per spectrum, scored as in Table~{V}. The two published rows are not molecule-disjoint and do not receive the same inputs as \PineIR.}
\label{tab:published_checkpoints}
\footnotesize
\begin{tabular}{llrrrr}
\toprule
Model & Training & Correct SMILES & Correct SMILES & Matched & Chemically \\
      & data     & on first try (\%) & within 50 tries (\%) & RMSD (\AA) & stable molecules (\%) \\
\midrule
DiffSpectra, published (IR + atom count) & QM9S & 35 & 58 & 0.31 & 72 \\
IR-GeoDiff, published (IR + formula)     & QM9S & 40 & 81 & 0.49 & 92 \\
\addlinespace
\PineIR (IR) & \PineDB $\rightarrow$ \QMeSDFT & 50 & 66 & 0.36 & 86 \\
\bottomrule
\end{tabular}
\end{table}

\clearpage
\section{Architecture and training objectives}
\label{sec:architecture}

\subsection{Stage 1: atom-latent autoencoder}
\label{sec:vae}

The encoder reads a 3D structure and returns one Gaussian posterior per atom, as in a variational autoencoder~\cite{kingmaAutoEncodingVariationalBayes2014}:
\begin{equation}
\mathbf{h}_i = \mathrm{Transformer}\bigl(\mathrm{MPNN}(a,\, r_{ij},\, \cos\theta)\bigr)_i, \qquad
\mathbf{z}_i = \boldsymbol{\mu}_i + \boldsymbol{\sigma}_i \odot \boldsymbol{\epsilon}_i, \quad \boldsymbol{\epsilon}_i \sim \mathcal{N}(0, I) .
\end{equation}
Here $a$ collects the atomic numbers of the molecule, $r_{ij}$ its interatomic distances and $\theta$ its bond angles; $\mathrm{MPNN}$ is a stack of six message-passing blocks that update edges from the angles they form and atoms from their edges, with distances expanded on radial basis functions~\cite{schuttSchNetDeepLearning2018, gasteigerDirectionalMessagePassing2022}. A Laplacian positional encoding of the molecular graph is added to the atom embeddings~\cite{dwivediGeneralizationTransformerNetworks2021}. $\mathrm{Transformer}$ is one attention layer over the atoms of the molecule~\cite{NIPS2017_3f5ee243}; $\mathbf{h}_i$ is the encoder feature of atom $i$; $\boldsymbol{\mu}_i$ and $\log\boldsymbol{\sigma}_i^2$ are linear in $\mathbf{h}_i$; $\boldsymbol{\epsilon}_i$ is Gaussian noise, $\odot$ the elementwise product, and $\mathbf{z}_i \in \mathbb{R}^{16}$ the latent of atom $i$. A molecule is therefore an unordered set of $N$ invariant latents. Unlike GeoLDM~\cite{xu2023geoldm}, which pairs invariant with equivariant latents, the coordinates are reconstructed from invariant latents rather than carried in equivariant latent vectors, closer to the unified latent space of UAE-3D~\cite{luoUnifiedLosslessLatent2025}.

The structure decoder first predicts the elements and a neighbour graph from the latents alone, trained to reproduce each atom's up to 12 nearest atoms within 5~\AA\ (the encoder's graph),
\begin{equation}
\eta_{ij} = \mathrm{MLP}\bigl([\,\mathbf{h}^\text{dec}_i + \mathbf{h}^\text{dec}_j,\; \lvert \mathbf{h}^\text{dec}_i - \mathbf{h}^\text{dec}_j \rvert\,]\bigr), \qquad
P_{ij} = \sigma\!\left(\tfrac{1}{2}(\eta_{ij} + \eta_{ji})\right),
\end{equation}
where $\mathbf{h}^\text{dec}_i$ is a linear projection of $\mathbf{z}_i$, the brackets concatenate, $\eta_{ij}$ is the logit of the pair, $\sigma$ the logistic function and $P_{ij}$ the probability that $j$ is a neighbour of $i$. At inference each atom keeps its 12 highest-scoring pairs with $P_{ij}>0.5$; no distance cutoff is applied. Coordinates are then produced by ten equivariant layers that start from random positions and move every atom along the directions to its predicted neighbours,
\begin{equation}
\mathbf{x}_i^{\,l+1} = \mathbf{x}_i^{\,l} + \frac{1}{\lvert \mathcal{N}(i) \rvert} \sum_{j \in \mathcal{N}(i)}
\tanh\!\Bigl(\phi\bigl(\mathbf{g}_i^{\,l}, \mathbf{g}_j^{\,l}, \mathbf{e}_{ij}^{\,l}\bigr)\Bigr)\,
\frac{\mathbf{x}_j^{\,l} - \mathbf{x}_i^{\,l}}{\sqrt{\lVert \mathbf{x}_j^{\,l} - \mathbf{x}_i^{\,l} \rVert^2 + \varepsilon^2}} ,
\end{equation}
with $\mathbf{x}^{\,l}_i$ the position of atom $i$ after layer $l$, $\mathcal{N}(i)$ its neighbours in the predicted graph, $\mathbf{g}^{\,l}_i$ and $\mathbf{e}^{\,l}_{ij}$ the atom and edge features of the layer, recomputed from the current geometry and re-injected with the latent, $\phi$ a small network and $\varepsilon$ a constant that keeps the direction finite. Only scalars multiply relative positions, so the update is E(3)-equivariant as in EGNN~\cite{satorrasEquivariantGraphNeural2021} and EDM~\cite{pmlr-v162-hoogeboom22a}. Unlike a diffusion decoder, the flow has no noise schedule and no time variable: the latent, not the number of steps, carries the geometry.

The \pIR decoder maps one latent at a time to a spectrum,
\begin{equation}
\widehat{\mathrm{pIR}}_i = \mathrm{softplus}\bigl(\mathrm{Deconv}(\mathbf{z}_i)\bigr) \in \mathbb{R}_{\ge 0}^{3500}, \qquad
\hat{S} = \sum_{i=1}^{N} \widehat{\mathrm{pIR}}_i ,
\end{equation}
where $\mathrm{Deconv}$ is a stack of transposed 1D convolutions and $\hat{S}$ the molecular IR implied by the latents. An atom's decoded \pIR depends on its own latent only. The per-molecule normalisation of Section~{II\,A} of the main text is written into the latent by the encoder, which sees the whole molecule, so the decoder needs neither the atom count nor the other atoms. The partial sums over an incomplete molecule therefore grow towards the known total area, which is what lets stage 2 use the decoder as a critic of incomplete molecules.

The autoencoder minimises
\begin{equation}
\mathcal{L}_1 = \mathcal{L}_\text{struct} + \lambda_\text{KL}\, D_\text{KL}\bigl(q(\mathbf{z} \mid G) \,\|\, \mathcal{N}(0, I)\bigr)
+ \lambda_\text{pIR}\, \mathcal{L}_\text{pIR}\bigl(\widehat{\mathrm{pIR}}, \mathrm{pIR}\bigr) + \lambda_\text{IR}\, \mathcal{L}_\text{IR}\bigl(\hat{S}, S\bigr) + \mathcal{L}_\text{gen},
\end{equation}
where
\begin{itemize}
\item $\mathcal{L}_\text{struct}$ sums a cross-entropy on the elements, a binary cross-entropy on the neighbour graph, a coordinate error taken after Kabsch alignment~\cite{kabschSolutionBestRotation1976} and penalties on local distances and atomic clashes;
\item $q(\mathbf{z} \mid G)$ is the encoder posterior and $D_\text{KL}$ the Kullback--Leibler divergence;
\item $\mathcal{L}_\text{pIR}$ combines a per-bin smooth-L1 loss, a 1-Wasserstein distance, a cosine distance and a relative area error on each atom's \pIR;
\item $\mathcal{L}_\text{IR}$ combines a per-bin smooth-L1 loss, a 1-Wasserstein distance and an intensity-weighted error on the molecular spectrum;
\item $\mathcal{L}_\text{gen}$ asks \pIR{}s decoded from prior samples $\mathbf{z} \sim \mathcal{N}(0, I)$ to have the area of real ones, and pulls the per-dimension mean and spread of the atom latents over a batch towards those of $\mathcal{N}(0, I)$, so that the latents sampled in stage~2 decode to plausible atoms;
\item the $\lambda$ are weights.
\end{itemize}
A small $\lambda_\text{KL}$ ($10^{-4}$) keeps the latents close to deterministic while bounding their scale, as in latent diffusion~\cite{rombachHighResolutionImageSynthesis2022}. The autoencoder is first pretrained on \PCQM molecules alone, with every spectral term off, and then trained with the structure and spectral objective on \PineDB; $\mathcal{L}_\text{gen}$ and the shape terms of $\mathcal{L}_\text{pIR}$ are switched on last, in two short fine-tunes.

\subsection{Stage 2: autoregressive latent generator}
\label{sec:ar}

The generator is trained against the frozen autoencoder and factorises the distribution over atoms:
\begin{equation}
p_\theta(\mathbf{z}_{1:N}, a_{1:N} \mid S) = P\bigl(\texttt{STOP} \mid S, \mathbf{z}_{\le N}\bigr) \prod_{i=1}^{N} P\bigl(a_i \mid S, \mathbf{z}_{<i}, R_{i-1}\bigr)\; p_\theta\bigl(\mathbf{z}_i \mid a_i, S, \mathbf{z}_{<i}, R_{i-1}\bigr),
\end{equation}
\begin{equation}
R_i = S - \sum_{j \le i} \widehat{\mathrm{pIR}}_j .
\end{equation}
Here $a_i$ is the element of atom $i$, $\mathbf{z}_{<i}$ the latents generated before it, and $R_i$ the residual spectrum, the part of $S$ that the first $i$ atoms do not yet explain, with $\widehat{\mathrm{pIR}}_j$ decoded by the frozen \pIR decoder. The number of atoms is not an input: generation ends when the element head predicts \texttt{STOP}. Training molecules are ordered by one fixed depth-first walk over the heavy atoms in canonical rank~\cite{landrumRDKitOpenSource}, with hydrogens on heteroatoms after their parent and those on carbon at the end, as in KRONOS~\cite{ottomano2026autoregressive}, so all conformers of a molecule share the order; local orders were shown to matter for 3D molecules~\cite{chengScalableAutoregressive3D2025, roseNEATNeighborhoodGuidedEfficient2026}.

\paragraph{Backbone.} A strided 1D convolution cuts the spectrum into 110 overlapping patch tokens with learned positions, as in vision transformers~\cite{dosovitskiyImageWorth16x162021} and in earlier IR models~\cite{wuTransformerBasedModelsPredicting2025, wangDiffSpectraMolecularStructure2025v2}. They form a prefix that every step attends to~\cite{raffelExploringLimitsTransfer2020}, followed by one token per step,
\begin{equation}
\mathbf{u}_i = W_z \mathbf{z}_{i-1} + \mathbf{E}_{a_{i-1}} + \rho(W_R R_{i-1}), \qquad
\mathbf{c}_i = \mathrm{Transformer}\bigl([\,\mathbf{s}_{1:110};\, \mathbf{u}_1, \dots, \mathbf{u}_i\,]\bigr)_i ,
\end{equation}
where $W_z$ projects the previous latent, $\mathbf{E}_a$ is the embedding of element $a$, $W_R$ a linear projection of the residual and $\rho$ a small network on it, $\mathbf{s}_{1:110}$ the spectrum tokens, $\mathbf{c}_i$ the context of step $i$, and a learned token replaces $\mathbf{u}_1$.  The causal transformer has sixteen layers of width 1024, and the step tokens carry no positional embedding, since the latents describe atomic environments rather than positions in a string. Because $W_R$ is linear, $W_R R_i = W_R S - \sum_{j \le i} W_R \widehat{\mathrm{pIR}}_j$: each \pIR is projected once and the residual is kept at full spectral resolution at the cost of a cumulative sum.

\paragraph{Element first, then latent.} A categorical head $P(a_i \mid \cdot) = \mathrm{softmax}(W_e \mathbf{c}_i)$ over the elements plus a \texttt{STOP} class chooses the element, with $W_e$ its projection. The latent is then drawn by a small diffusion head conditioned on that choice, following MAR~\cite{li2024autoregressive} and KRONOS~\cite{ottomano2026autoregressive}. The head is a residual MLP with adaptive layer normalisation~\cite{perezFiLMVisualReasoning2018, peeblesScalableDiffusionModels2023} conditioned on $\mathbf{c}^\text{head}_i = [\,\mathbf{c}_i + \mathbf{E}_{a_i};\, \bar{\mathbf{s}}\,]$, with $\bar{\mathbf{s}}$ the mean of the spectrum tokens, and is trained by noise prediction~\cite{NEURIPS2020_4c5bcfec} on a cosine schedule~\cite{pmlr-v139-nichol21a}:
\begin{equation}
\mathbf{z}_{i,t} = \sqrt{\bar\alpha_t}\, \mathbf{z}_i + \sqrt{1 - \bar\alpha_t}\, \boldsymbol{\epsilon}, \qquad
\mathcal{L}_2 = \mathbb{E}_{t, \boldsymbol{\epsilon}} \bigl\lVert \boldsymbol{\epsilon}_\theta(\mathbf{z}_{i,t}, t, \mathbf{c}^\text{head}_i) - \boldsymbol{\epsilon} \bigr\rVert^2 + \lambda_\text{elem}\, \mathrm{CE}\bigl(P(a_i), a_i^\star\bigr) .
\end{equation}
Here $t \in [0, 1]$ is the diffusion time, $\bar\alpha_t$ the schedule, $\mathbf{z}_{i,t}$ the target latent noised to time $t$, $\boldsymbol{\epsilon}_\theta$ the head's noise prediction, $a_i^\star$ the reference element (\texttt{STOP} at step $N+1$) and $\lambda_\text{elem}$ a weight. The targets are the posterior means $\boldsymbol{\mu}_i$, standardised per dimension. To reduce exposure bias~\cite{NIPS2015_e995f98d}, we perturb a fifth of the latents fed back as inputs with Gaussian noise, as in NEAT~\cite{roseNEATNeighborhoodGuidedEfficient2026}, and compute the residual from the perturbed latents.

During training, the encoded spectrum is replaced by learned null tokens with probability 0.1 per molecule, following the condition-dropout strategy used for classifier-free guidance~\cite{hoClassifierFreeDiffusionGuidance2022}. The residual spectrum and preceding atom latents remain available. At inference, we use the conditional prediction directly, without classifier-free guidance.

Ordering the element before the latent also makes an extra input a sampling-time option: masking \texttt{STOP} until $N$ atoms gives an IR + atom-count model, and masking elements once they are used up gives an IR + formula model. This allows input-matched comparisons with DiffSpectra~\cite{wangDiffSpectraMolecularStructure2025v2} and IR-GeoDiff~\cite{wuLatentDiffusionbased3D2026}. The main results use \PineIR with neither extra input. Table~{III} of the main text and Tables~\ref{tab:full_external} and~\ref{tab:full_indist} report both input-matched variants.

\clearpage
\section{Checkpoints and result provenance}
\label{sec:checkpoints}

The main-text model names refer to the checkpoints in Table~\ref{tab:checkpoints}. Tables~{III}--{IV} of the main text use models trained on \PineDB at the SO3LR level only; Table~{V} uses models fine-tuned on, or trained only on, \QMeSDFT. The checkpoint table separates model identity from the evaluation protocol.

\begin{table}[htb!]
\caption{Model versions, training data and starting components. Rows 1--4 are the \PineDB models of Tables~{III}--{IV}; rows 5--8 are the models of Table~{V}. \QMeSSO evaluations are external, while \PineDB evaluations use its own held-out distribution.}
\label{tab:checkpoints}
\footnotesize
\renewcommand{\arraystretch}{1.2}
\begin{tabular}{p{3.8cm}p{4.0cm}p{4.2cm}p{4.2cm}}
\toprule
Model & Training data & Start model & Purpose \\
\midrule
\PineIR (atom-wise latent generator) & \PCQM + \GEOM SO3LR, joint \PineDB split; no DFT stage & Frozen stage-1 autoencoder of row 4 & Generate variable-size structures and rank candidates by their predicted spectra. \\
\addlinespace
DiffSpectra (3D diffusion baseline) & Same joint \PineDB SO3LR split; no DFT stage & Continued from an earlier run with a restarted learning rate; pretrained spectrum encoder & Compare generation when the atom count is supplied. \\
\addlinespace
IR-GeoDiff (latent diffusion baseline) & Same joint \PineDB SO3LR split; no DFT stage & Diffusion stage continued from an earlier run, on a retrained autoencoder; frozen classifier and autoencoder & Compare generation when the molecular formula is supplied. \\
\addlinespace
Stage-1 autoencoder (structure and spectrum reconstruction) & Structure-only \PCQM pretraining, then joint \PineDB SO3LR structure/spectrum training & Structure encoder and decoder before adding spectral objectives & Supply \PineIR's latent representation and evaluate atom-wise forward predictions. \\
\addlinespace
\PineIR, fine-tuned & \PineDB SO3LR, then \QMeSDFT train & \PineIR of row 1; frozen autoencoder of row 4 with its \pIR decoder fine-tuned on DFT & Transfer the pretrained generator to DFT spectra. \\
\addlinespace
\PineIR, \QMeSDFT only & \QMeSDFT train only & Generator from scratch; frozen autoencoder trained on \QMeSDFT after structure-only \PCQM pretraining & Measure what \PineDB pretraining contributes. \\
\addlinespace
DiffSpectra, fine-tuned & \PineDB SO3LR, then \QMeSDFT train & DiffSpectra of row 2 with a distilled spectrum encoder & DFT baseline that receives the atom count. \\
\addlinespace
DiffSpectra and IR-GeoDiff, published & QM9S (DFT) & As released & Literature reference on the HCNOF subset, Section~\ref{sec:published_checkpoints} (Table~\ref{tab:published_checkpoints}). \\
\bottomrule
\end{tabular}
\end{table}

\clearpage
\section{References}
\bibliographystyle{achemso}
\bibliography{library}